\documentclass[twocolumn,pra,superscriptaddress,nofootinbib]{revtex4-1}
\usepackage{amssymb}
\usepackage{times}
\usepackage{color}
\usepackage{graphicx}
\usepackage[dvipsnames]{xcolor}
\usepackage{physics}
\usepackage{bm}
\usepackage{mathtools}
\usepackage{upgreek}
\usepackage[colorlinks=true,allcolors=blue]{hyperref}
\usepackage{soul}
\usepackage{footmisc}

\begin{document}
\title{Circuit quantization in the presence of time-dependent external flux}

\author{Xinyuan You}
\affiliation{Northwestern--Fermilab Center for Applied Physics and Superconducting Technologies, 
             Northwestern University, Evanston, Illinois 60208, USA}
\affiliation{Graduate Program in Applied Physics, Northwestern University, Evanston, Illinois 60208, USA}
\author{J.~A.~Sauls}
\affiliation{Northwestern--Fermilab Center for Applied Physics and Superconducting Technologies, 
             Northwestern University, Evanston, Illinois 60208, USA}
\affiliation{Department of Physics and Astronomy, Northwestern University, Evanston, Illinois 60208, USA}
\author{Jens Koch}
\affiliation{Northwestern--Fermilab Center for Applied Physics and Superconducting Technologies, 
             Northwestern University, Evanston, Illinois 60208, USA}
\affiliation{Department of Physics and Astronomy, Northwestern University, Evanston, Illinois 60208, USA}

\begin{abstract}
Circuit quantization links a physical circuit to its corresponding quantum Hamiltonian. The standard quantization procedure generally assumes any external magnetic flux to be static. Time dependence naturally arises, however, when flux is modulated or when flux noise is considered. In this case, application of the existing quantization procedure can lead to inconsistencies. To resolve these, we generalize circuit quantization to incorporate time-dependent external flux.
\end{abstract}
\maketitle

\section{Introduction} 
\label{sec:introduction}
Superconducting circuits are electrical circuits fabricated from superconducting materials. Due to the flexibility in circuit design, such circuits hold substantial promise as qubits~\cite{Devoret2013}, exhibiting coherence times now approaching the millisecond scale~\cite{Nguyen2018}. The link between a physical circuit and its quantum mechanical behavior is provided by circuit quantization~\cite{qnetwork,devoret1995quantum,Burkard2004,Nigg2012,Ulrich2016,vool,Parra-Rodriguez2019}. This implementation of canonical quantization has been widely used in the context of superconducting qubits and circuit quantum electrodynamics (cQED), but is restricted to circuits enclosing time-independent external magnetic flux\footnote{Time-dependent flux is mentioned by Burkard \textit{et al.}\ \cite{Burkard2004}, but kinetic terms
$\propto\dot{\Phi}_\text{e}$ are not discussed in that work.}. Recently, there have been a number of studies utilizing time-varying external flux. Alternating-current flux modulation has been proposed for realizing fast two-qubit gates via first-order sideband transitions~\cite{Beaudoin2012,Strand2013}. Parametric flux modulation also enables random access multi-qubit control, thus improving qubit connectivity in quantum processors~\cite{Naik2017,Reagor2018,Didier2018}. Furthermore, modulating the coupling strength between a qubit and a resonator by means of a time-dependent flux threading a superconducting quantum interference device (SQUID)  allows for universal stabilization of arbitrary single-qubit states~\cite{Lu2017a,huang2018}. In addition to parametric modulation, time-dependence is also crucial for the modeling of flux-induced qubit dissipation and dephasing~\cite{Wellstood1987,Ithier2005,Yoshihara2006,Kumar2016}. Therefore, it is timely to examine the general framework of circuit quantization for time-varying external flux.

In standard circuit quantization there exists a gauge freedom when formulating the Hamiltonian for a circuit threaded by external flux. Depending on the choice of gauge, the external flux may be associated with any of the potential energy terms. For example, in a fluxonium qubit~\cite{Manucharyan2009} the external flux may be associated with the potential energy of the Josephson junction \cite{Spilla2015,Sete2017,Nguyen2018}, or that of the superinductor \cite{Diggins1997,Koch2009}. However, this freedom leads to inconsistent predictions of the rate of qubit relaxation induced by fluctuations of the external flux, as we show below. Our generalization of circuit quantization to include time-dependent external flux resolves these inconsistencies and provides a formulation capable of handling a wider variety of circuits.

The paper is organized as follows. In Sec.~\ref{sec:time_independent}, we review the standard procedure of circuit quantization that is valid for time-independent external flux. We show in Sec.~\ref{sec:inconsistent} that inconsistencies arise if this procedure is applied to circuits coupled to time-varying flux; we specifically analyze qubit relaxation due to flux noise. In Sec.~\ref{sec:time_dependent}, we generalize the circuit-quantization framework for two concrete examples: the dc SQUID circuit and the fluxonium qubit with time-dependent external flux. The concept of irrotational degrees of freedom is introduced, and the previous inconsistency in the relaxation time is resolved. 
Canonical quantization for general single-loop and multi-loop circuits with time-dependent flux is presented in Sec.~\ref{sec:general}. We summarize our work in Sec.~\ref{sec:summary}.

\section{Circuit quantization with time-independent external flux}\label{sec:time_independent}
\begin{figure}
\includegraphics[width=0.38\textwidth]{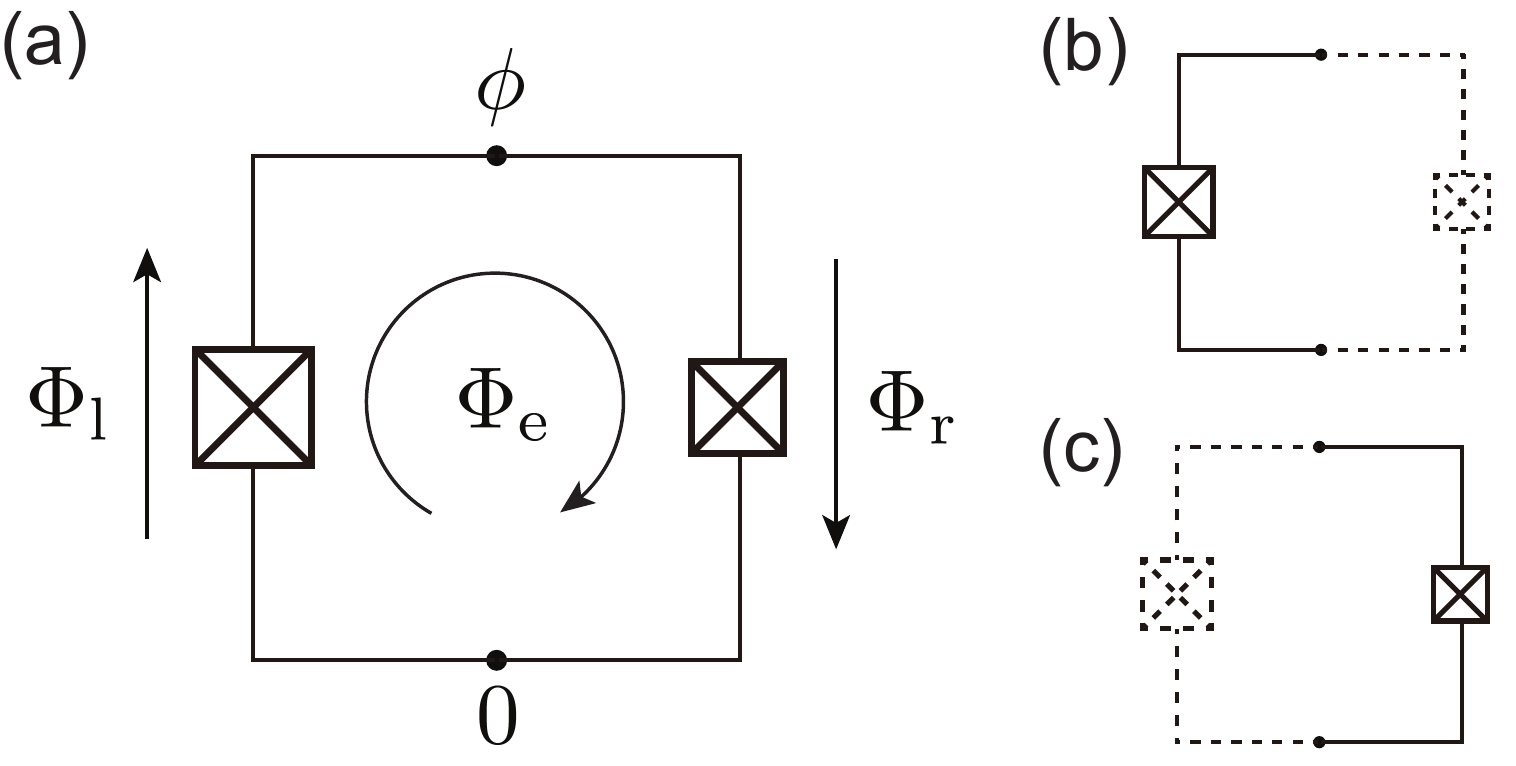}
\caption{\label{fig:squid} (a) Asymmetric SQUID formed by two Josephson junctions. $\Phi_\text{l}$ and $\Phi_\text{r}$ are branch flux variables for the left and right junction. The node flux $\phi$ is associated with the upper node, while the lower node is chosen to be the ground node. An external flux $\Phi_\text{e}$ threads the loop. (b) and (c) depict the two equivalent choices of spanning trees (solid lines) and closure branches (dashed lines).}
\end{figure}

We begin with a brief review of standard circuit quantization, valid for time-independent flux. The starting point is the construction of the Lagrangian for the circuit. Kinetic energy terms in this Lagrangian correspond to capacitive energies. The potential energy is composed of the energies of all inductive elements, including those associated with Josephson junctions. As a concrete example, we apply circuit quantization to an asymmetric dc SQUID consisting of two Josephson junctions with junction capacitances $C_\text{l}$, $C_\text{r}$, and Josephson energies $E_\text{Jl}$, $E_\text{Jr}$ [Fig.~\ref{fig:squid}(a)]. The static external flux, $\Phi_\text{e}$, which threads the circuit loop, is treated as a classical variable.

Each branch flux variable is defined as the time integral of the voltage across the corresponding circuit element. The dc SQUID has two branch variables, $\Phi_\text{l}$ and $\Phi_\text{r}$, for the left and right arm of the SQUID, respectively. Following Ref.~\onlinecite{devoret1995quantum}, we also define flux variables associated with each circuit node. Setting the lower node in Fig.~\ref{fig:squid}(a) to be the ground node, the system can be described by one active node flux $\phi$ corresponding to the upper node. If we choose the left arm to form the spanning tree, then the right arm becomes the closure branch [see Fig.~\ref{fig:squid}(b)]. By convention, one allocates the external flux to the closure branch \cite{devoret1995quantum}. This step guarantees fluxoid quantization for time-independent external flux \cite{barone1982physics,orlando1991foundations}. The relation between the node variable $\phi$ and the branch variables is thus
\begin{equation}\label{node_l}
\Phi_\text{l}=\phi\,, \qquad \Phi_\text{r}=-\phi+\Phi_\text{e}\,.
\end{equation}

In terms of the node flux, the Lagrangian can be written as a function of one independent degree of freedom,
\begin{equation}
\mathcal{L}_\text{l} = \dfrac{1}{2} C_\Sigma \dot{\phi}^2 
                     + E_\text{Jl}\cos{\dfrac{2\pi}{\Phi_0}\phi} 
                     + E_\text{Jr}\cos{\dfrac{2\pi}{\Phi_0}(-\phi+\Phi_\text{e})}\,,
\end{equation}
where $C_\Sigma=C_\text{l}+C_\text{r}$ is the total capacitance, and $\Phi_0=h/2e$ the flux quantum. The Hamiltonian is obtained by performing a Legendre transformation with generalized momentum $Q = \partial \mathcal{L}_\text{l}/\partial \dot{\phi} = C_\Sigma \dot{\phi}$. Employing canonical quantization in the coordinate representation, $Q=-i\hbar\,d/d\phi \equiv 2en$, and denoting $E_\text{C} = e^2/2C_\Sigma$, the Hamiltonian takes the form 
\begin{equation}\label{inleft}
\mathcal{H}_\text{l} = 4E_\text{C}n^2 
                     - E_\text{Jl}\cos{\dfrac{2\pi}{\Phi_0}\phi} 
                     - E_\text{Jr}\cos{\dfrac{2\pi}{\Phi_0}(\phi-\Phi_\text{e})}
\,.
\end{equation}

In Eq.~\eqref{inleft} we chose the right arm to be the closure branch. Equivalently, we may choose the left arm as the closure branch [Fig.~\ref{fig:squid}(c)]. This choice groups the external flux with the left arm. The two branch fluxes then take on the form
\begin{equation}\label{node_r}
\Phi_\text{l}=\phi + \Phi_\text{e}\,, \qquad \Phi_\text{r}=-\phi\,,
\end{equation}
resulting in the Hamiltonian 
\begin{equation}\label{inright}
\mathcal{H}_\text{r} = 4E_\text{C}n^2 - E_\text{Jl} \cos{\dfrac{2\pi}{\Phi_0}(\phi+\Phi_\text{e})} - E_\text{Jr}\cos{\dfrac{2\pi}{\Phi_0}\phi}\,.
\end{equation}
The two Hamiltonians $\mathcal{H}_\text{l}$ and $\mathcal{H}_\text{r}$ are related by the unitary transformation
\begin{equation}\label{transformation1}
\mathcal{H}_\text{r} = U^\dag \mathcal{H}_\text{l} U\,, \qquad U=\exp(-\dfrac{i}{\hbar}Q\Phi_\text{e})
\end{equation}
which implies they have the same eigenvalue spectrum, and two sets of eigenfunctions shifted by $\Phi_\text{e}$ in the variable $\phi$.

\section{Inconsistent predictions of qubit relaxation times}\label{sec:inconsistent}
While the two Hamiltonians obtained for different choices of closure branches predict equivalent results for static flux, inconsistencies arise if we directly apply standard quantization to study qubit relaxation caused by a fluctuating external flux. We will trace back this issue to the proper treatment of the time-dependent holonomic constraint imposed by fluxoid quantization in Sec.~{\ref{sec:time_dependent}}. Suppose the flux is randomly varying around some fixed, but controllable, value $\Phi_\text{e}^0$,
\begin{equation}\label{noise}
\Phi_\text{e}(t) = \Phi_\text{e}^0 + \delta \Phi_\text{e}(t)\,.
\end{equation}
In general fluctuations induce transitions between the qubit states $|g\rangle$ and $|e\rangle$. According to Fermi's golden rule, the decay rate of the qubit state $\ket{e}$ is proportional to~\cite{Ithier2005} 
\begin{equation}\label{fermi}
T_1^{-1} \propto |\langle g| \partial_{\Phi_\text{e}} \mathcal{H}   | e\rangle|^2\,,
\end{equation}
where the partial derivative is evaluated at $\Phi_\text{e} (t) = \Phi_\text{e}^0$. Equation~\eqref{fermi} implies that the qubit relaxation rate would depend on the choice of closure branch because of the association of the external flux with different terms in the Hamiltonian. In particular, the Hamiltonians $\mathcal{H}_\text{l}$ and $\mathcal{H}_\text{r}$ exhibit two different instances of flux grouping: The former allocates flux to the right branch, and the resulting relaxation rate would be proportional to $E_\text{Jr}^2$. In the latter case, flux is grouped with the left branch, and the relaxation rate would be proportional to $E_\text{Jl}^2$. In an asymmetric SQUID the two Josephson energies differ, $E_\text{Jl}\neq E_\text{Jr}$, and thus the qubit relaxation rate seemingly depends on how we allocate the flux in the Hamiltonian\footnote{We note that the switch from eigenstates of $\mathcal{H}_\text{r}$ to those of $\mathcal{H}_\text{l}$ does not compensate the observed different scaling with $E_\text{Jl}$ and $E_\text{Jr}$, respectively.}. This inconsistency is not limited to an asymmetric SQUID loop, but generally applies to circuit loops with distinct potential energy coefficients. For example, the external flux in a fluxonium qubit~\cite{Manucharyan2009} can be allocated either to the inductor~\cite{Diggins1997,Koch2009} or the Josephson junction~\cite{Spilla2015,Sete2017,Nguyen2018} which would lead to different predictions for the relaxation time due to flux noise\footnote{Typically, relaxation due to flux noise is subdominant due to the 1/$f$ nature of the noise, so that conclusions in the literature are not changed to the best of our knowledge.}. We emphasize that the transformation {\eqref{transformation1}} is merely a variable displacement, and hence unrelated to rotating-frame transformations that may affect decoherence rates~{\cite{Brown2007,Jing2014}}. This inconsistency rather hinges upon the correct treatment of time-dependent fluxoid quantization, as we will see next.

\section{Circuit quantization for time-dependent external flux: dc SQUID and Fluxonium}\label{sec:time_dependent}
We next formulate circuit quantization in the presence of time-dependent flux for the concrete examples of the dc SQUID circuit and the fluxonium qubit. The general discussion of arbitrary circuits is presented in the subsequent section.

\subsection{From fluxoid quantization to the time-dependent Hamiltonian: dc SQUID}\label{sec-DC-SQUID} 
We begin with the constraint imposed by fluxoid quantization. Most naturally, it is written in terms of branch variables,
\begin{equation}\label{fluxoid}
\Phi_\text{l} + \Phi_\text{r} = \Phi_\text{e}(t)
\,,
\end{equation}
which is a time-dependent holonomic constraint. According to it, the sum of the branch variables is fixed by the external flux, and hence is not a dynamical variable. The single degree of freedom, denoted by the variable $\widetilde{\Phi}$, can be expressed as a linear combination of the two branch variables,
\begin{equation}\label{dofvar}
\widetilde{\Phi} = m_\text{l} \Phi_\text{l} + m_\text{r} \Phi_\text{r}
\,,
\end{equation}
where $m_\text{l} \neq m_\text{r}$. The node variables defined in  Eqs.~{\eqref{node_l}} and {\eqref{node_r}} are obtained by setting $(m_\text{l},m_\text{r})=(1,0)$ and $(0,-1)$, respectively.

To derive the time-dependent Hamiltonian, we first express the circuit Lagrangian in terms of the two branch variables,
\begin{equation}\label{lagbranch}
\mathcal{L} = \dfrac{1}{2}C_\text{l}\dot{\Phi}_\text{l}^2 
            + \dfrac{1}{2}C_\text{r}\dot{\Phi}_\text{r}^2 
            + E_\text{Jl}\cos{\varphi_\text{l}} 
            + E_\text{Jr}\cos{\varphi_\text{r}}
\,,
\end{equation}
where all lower-case $\varphi$'s denote reduced flux variables, $\varphi=2\pi \Phi/\Phi_0$. Next, we transform
to the variable {$\tilde\Phi$} and the external flux {$\Phi_\text{e}$} by inverting Eqs.~\eqref{fluxoid} and \eqref{dofvar}. This results in
\begin{equation}\label{branch_transformations}
{
	\Phi_\text{l} = (\widetilde{\Phi} - m_\text{r}\Phi_\text{e})/m_\Delta\,,\qquad
	\Phi_\text{r} = -(\widetilde{\Phi} - m_\text{l}\Phi_\text{e})/m_\Delta\,,
}
\end{equation}
where $m_\Delta=m_\text{l}-m_\text{r}$. 
We now see that proper implementation of the time-dependent holonomic constraint~{\eqref{fluxoid}} yields
\begin{align}
\label{lag}
\mathcal{L}(m_\text{l},m_\text{r}) 
= 
& \dfrac{C_\Sigma}{2m_\Delta^2}\dot{\widetilde{\Phi}}^2 - \dfrac{C_\text{r}m_\text{l}+C_\text{l}m_\text{r}}{m_\Delta^2}\dot{\widetilde{\Phi}}\dot{\Phi}_\text{e} 
\\
& + E_\text{Jl}\cos{\dfrac{\widetilde{\varphi}-m_\text{r}\varphi_\text{e}}{m_\Delta}} + E_\text{Jr}\cos{\dfrac{\widetilde{\varphi}-m_\text{l}\varphi_\text{e}}{m_\Delta}}\,,
\nonumber
\end{align}
where we dropped the term $\propto \dot{\Phi}_\text{e}^2$.
Performing a Legendre transformation leads to the Hamiltonian
\begin{align}
\label{ham}
\mathcal{H}(m_\text{l},m_\text{r}) = & 4E_\text{C}m_\Delta^2 n^2 + \dfrac{C_\text{r}m_\text{l}+C_\text{l}m_\text{r}}{C_\Sigma}2en\dot{\Phi}_\text{e}
\\
\nonumber
&-E_\text{Jl}\cos{\dfrac{\widetilde{\varphi}-m_\text{r}\varphi_\text{e}}{m_\Delta}} - E_\text{Jr}\cos{\dfrac{\widetilde{\varphi}-m_\text{l}\varphi_\text{e}}{m_\Delta}}\,. 
\end{align}
The term $\propto\dot{\Phi}_\text{e}$ is missing in standard circuit quantization for Hamiltonians parametrized only by the time-independent flux, $\Phi_\text{e}$.
For special choices of $(m_\text{l},m_\text{r})$ obeying\footnote{We note that there remains a continuous family of choices for $(m_\text{l},m_\text{r})$ satisfying Eq.~{\eqref{irrcondition}}. This freedom is characterized by the scaling factor $m_\Delta$. In Sec.~{\ref{sec:general}}, the remaining freedom is expressed by the matrix $\mathbf{A}$.\label{foot}}
\begin{equation}\label{irrcondition}
C_\text{r}m_\text{l}+C_\text{l}m_\text{r} = 0
\,,
\end{equation}
the linear coupling of the charge operator to the fluctuating EMF generated by $\dot\Phi_\text{e}$ vanishes.
We refer to Eq.~\eqref{irrcondition} as the \emph{irrotational} constraint\footnote{As the $\dot\Phi_\text{e}$ term is related to a time-dependent unitary transformation (rotating-frame transformation), we call the constraint eliminating this term ``irrotational".} on the variable $\widetilde{\Phi}$, and to $\widetilde{\Phi}$ with the constraint imposed as the irrotational operator, variable or degree of freedom. 

In general, Hamiltonians for different choices $(m_\text{l},m_\text{r})$ and $(m_\text{l}',m_\text{r}')$ are related by a time-dependent unitary transformation  $\mathcal{U}(t) = S\,U(t)$, consisting of a gauge transformation and a scale transformation. The gauge transformation is given by
\begin{equation}
U(t)=\exp(\dfrac{i}{\hbar}Q 
\dfrac{m_\text{l}m_\text{r}'-m_\text{l}'m_\text{r}}{m_\Delta}
\Phi_\text{e}(t))
\end{equation}
and the scaling transformation by
\begin{equation}
S = \exp(\dfrac{i}{2\hbar}\ln(\dfrac{m_\Delta'}{m_\Delta})G)
\,,
\end{equation}
where the generator is the anti-commutator of the conjugate coordinate and momentum, $G = \widetilde{\Phi}Q + Q\widetilde{\Phi}$. The action of $S$ on the conjugate variables is 
\begin{equation}
S^\dag Q S = \dfrac{m_\Delta'}{m_\Delta} Q\,,
\qquad 
S^\dag \widetilde{\Phi} S = \dfrac{m_\Delta}{m_\Delta'}\widetilde{\Phi}
\,. 
\end{equation}
Since operators and states transform according to
\begin{equation}
\mathcal{A}' = \mathcal{U}^\dag(t) \mathcal{A}\,\mathcal{U}(t)\,, \qquad |\psi'\rangle = \mathcal{U}^\dag(t) |\psi\rangle\,,
\end{equation}
all expectation values,  measured at one particular time, are invariant under the unitary transformation $\mathcal{U}(t)$:
$\langle \psi'| \mathcal{A}'|\psi'\rangle
=\langle \psi| \mathcal{A} | \psi \rangle$.
The Schr{\"o}dinger equation also remains form invariant, with the transformed Hamiltonian given by
\begin{equation}\label{Utransform}
\mathcal{H}(m_\text{l}',m_\text{r}')
=
\mathcal{U}^\dag(t)\mathcal{H}(m_\text{l},m_\text{r})\mathcal{U}(t)
-i\hbar\,\mathcal{U}^\dag(t)\partial_t \mathcal{U}(t)\,.
\end{equation}
However, as we discuss below, multi-time correlation functions are in general not invariant under the time-dependent unitary transformation.

\subsection{Multi-time observables\label{sec:correlator}}
Consider the probability 
\begin{equation}
C(0,t) = |\langle \psi(0)|\psi(t)\rangle|^2
\,,
\end{equation}
for transitions induced by flux noise. The decay rate of the transition probability averaged over noise realizations, $\langle C(0,t)\rangle_\text{av}$, is directly related to the spontaneous relaxation rate. For simplicity, we will ignore the usual 1/$f$ character of flux noise. Instead, we consider stationary Gaussian noise with short correlation time $t_\text{c}$, standard deviation $\sigma$, and noise power spectrum $S_{\Phi_\text{e}}(\omega)$, defined as the Fourier transform of the autocorrelation function. 

To calculate the noise-averaged transition probability, fluctuations in both $\Phi_\text{e}$ and $\dot{\Phi}_\text{e}$ must be taken into account. We prepare the system in the first excited eigenstate 
$|\psi(0)\rangle=|e^0(m_\text{l},m_\text{r})\rangle$ of the unperturbed Hamiltonian 
\begin{align}\label{ham0}
\mathcal{H}^0 (m_\text{l},m_\text{r}) = & 4E_\text{C}m_\Delta^2 n^2
\\
\nonumber
& - E_\text{Jl}\cos{\dfrac{\widetilde{\varphi}-m_\text{r}\varphi^0_\text{e}}{m_\Delta}} - E_\text{Jr}\cos{\dfrac{\widetilde{\varphi}-m_\text{l}\varphi^0_\text{e}}{m_\Delta}}\,,
\end{align}
obtained by omitting all fluctuating terms from $\mathcal{H}(m_\text{l},m_\text{r})$ in Eq.~{\eqref{ham}}. We proceed by evolving the initial state under the full Hamiltonian $\mathcal{H}(m_\text{l},m_\text{r})$ to some later time $t$, then carry out the average over noise realizations. As shown in Appendix~\ref{sec:correlation}, for $t_\text{c}\ll t\ll T_1$ one obtains the perturbative result
\begin{align}
\langle C(0,t)\rangle_\text{av} = & 1 -|\langle g^0_\text{irr} | \partial_{\Phi_\text{e}} 
\mathcal{H}_\text{irr} | e^0_\text{irr}\rangle|^2  S_{\Phi_\text{e}}(\omega_{\text{eg}})t/\hbar^2 
\nonumber
\\
&-\eta^2(m_\text{l}, m_\text{r})|\langle e^0_\text{irr} | n | g^0_\text{irr} \rangle|^2 8\sigma^2e^2/\hbar^2 
\,.
\label{corr}
\end{align}
Here, $\mathcal{H}^0_\text{irr}$ and $\mathcal{H}_\text{irr}$ are obtained from $\mathcal{H}^0$ [Eq.~{\eqref{ham0}}] and $\mathcal{H}$ [Eq.~{\eqref{ham}}] by satisfying the irrotational constraint~{\eqref{irrcondition}} with
\begin{equation}\label{irrot}
m_\text{l} = \dfrac{C_\text{l}}{C_\Sigma}\,,\qquad m_\text{r} = -\dfrac{C_\text{r}}{C_\Sigma}\,,
\end{equation}
where we take $m_\Delta=1$ for simplicity. $|g_\text{irr}^0\rangle$ and $|e_\text{irr}^0\rangle$ are the lowest two eigenstates of $\mathcal{H}^0_\text{irr}$, with energy difference $\omega_\text{eg}$. While Eq.~{\eqref{corr}} is expressed in terms of quantities with irrotational constraints, it is valid for general choice of $(m_\text{l},m_\text{r})$.

There are two key features to the result for the transition probability in Eq.~\eqref{corr}. First, the 
averaged transition  rate is
$\Gamma_1=|\langle g^0_\text{irr}|\partial_{\Phi_\text{e}}\mathcal{H}_\text{irr}|e^0_\text{irr}\rangle|^2 
S_{\Phi_\text{e}}(\omega_{\text{eg}})/\hbar^2$. 
Hence there is no ambiguity in the relaxation rate 
with respect to the choice of $(m_\text{l},m_\text{r})$.
Secondly, the calculation yields an offset term which depends on the choice of gauge. This unphysical offset is eliminated if we impose the irrotational constraint of Eq.~\eqref{irrcondition}.
A similar gauge-dependent term appears in the expression for pure dephasing of the qubit, as discussed in Appendix \ref{sec:pure_dephasing_time}. 
While the procedure outlined above provides a satisfactory theory for the transition rate, the gauge-fixing requirement used to eliminate the unphysical offset in the transition probability requires further analysis and discussion.

\begin{figure}
    \includegraphics[width=0.45\textwidth]{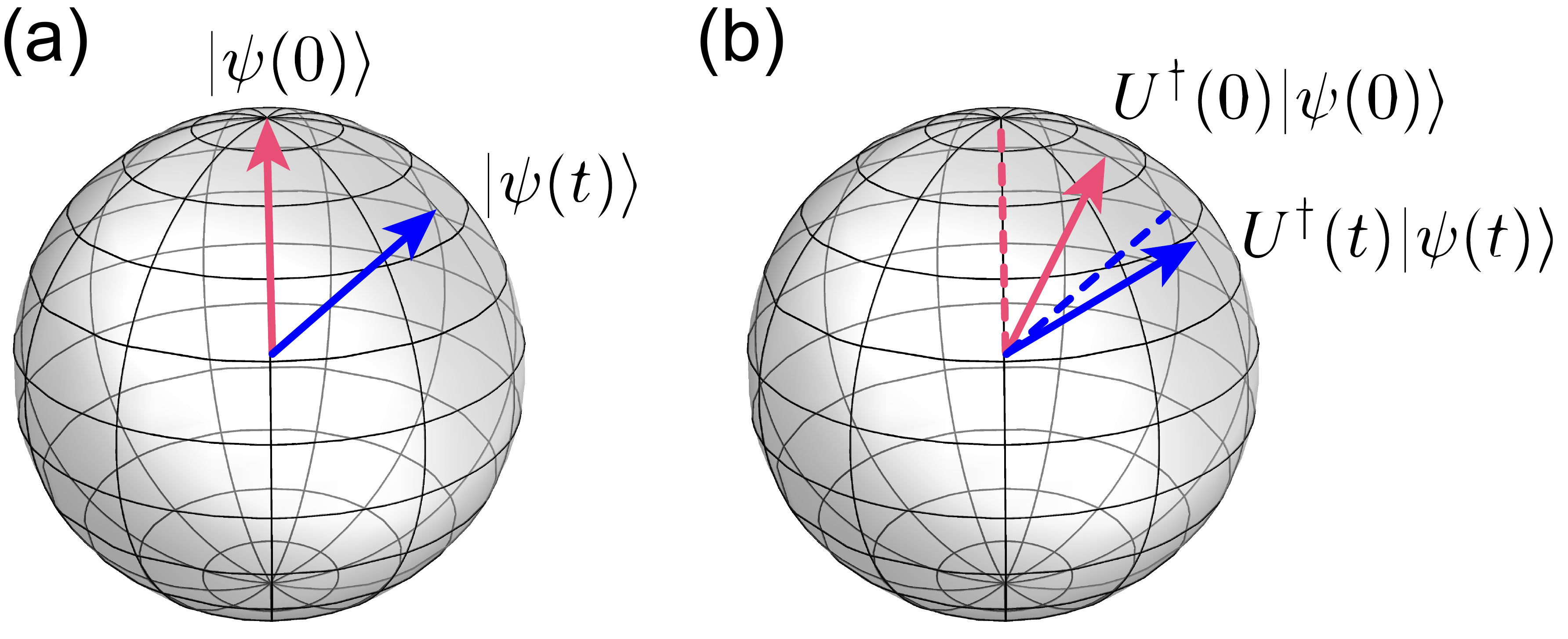}
    \caption{(color online) The overlap between two states at different times is not conserved under a time-dependent unitary transformation, since $\langle\psi(0)|U(0)U^\dag(t)|\psi(t)\rangle$$\neq$$\langle\psi(0)|\psi(t)\rangle$ in general.}
    \label{fig:correlator}
\end{figure}

The fact that time-dependent unitary transformations can affect the values of multi-time correlation functions is not surprising. For states defined as points on the Bloch sphere, Fig.~\ref{fig:correlator} illustrates how a time-dependent rotating-frame transformation, carried out at two different times, generally leads to different state-vector overlaps.
In the following we examine the origin of the unphysical offset term in the transition probability in Eq.~\eqref{corr}.

Consider the limit in which the system parameters $E_\text{C}$, $E_\text{Jl}$, $E_\text{Jr}$ tend to zero. Despite the fact that the system Hamiltonian should vanish in this limit,  
Eq.~\eqref{ham} leads to
\begin{equation}\label{Hlimit}
\mathcal{H}(m_\text{l},m_\text{r}) \longrightarrow 2e\,\overline{\eta}(m_\text{l},m_\text{r})n\dot{\Phi}_\text{e}\,, 
\end{equation}
where $\overline{\eta}(m_\text{l},m_\text{r})=\eta(m_\text{l},m_\text{r})m_\Delta$ in general remains finite when taking the limit with fixed capacitance ratios. 
A non-vanishing value of for $\bar\eta(m_{\text{l}},m_{\text{r}})$ not only conflicts with the vanishing of the system Hamiltonian, but implies a gauge-dependent term proportional to $\dot\Phi_\text{e}$ in the Hamiltonian.
One easily confirms that this term originates from the derivative term in the time-dependent unitary transformation, $-i\hbar\, \mathcal{U}^\dag \partial_t \mathcal{U}$, and that it vanishes when 
the irrotational constraint is imposed. Thus, the irrotational constraint eliminates the unphysical term, and renders the Hamiltonian form invariant.

Finally, we show that the spurious offset term in Eq.~\eqref{corr} is generated by the unphysical term in the Hamiltonian, Eq.~\eqref{Hlimit}. Consider the time evolution of the excited state generated by this Hamiltonian:
\begin{equation}
|\psi(t)\rangle = \exp[\dfrac{1}{i\hbar}\int_0^t dt' 2e\overline{\eta} n\dot{\Phi}_\text{e}(t')]
| e^0(m_\text{l},m_\text{r}) \rangle\,.
\end{equation}
Expanding to order $O(\delta \Phi_\text{e})$ in the fluctuations of $\Phi_{\text{e}}(t)$, the overlap with the ground state becomes
\begin{align}
&\langle g^0(m_\text{l},m_\text{r})|\psi(t)\rangle\\
&\qquad = \dfrac{2e}{i\hbar} \overline{\eta}\, \langle g^0(m_\text{l},m_\text{r}) 
         |n|e^0(m_\text{l},m_\text{r})\rangle [\delta\Phi_\text{e}(t) - \delta\Phi_\text{e}(0)] 
\nonumber 
\\
\nonumber
&\qquad  = \dfrac{2e}{i\hbar} \eta\,
\langle g^0_\text{irr} |n|e^0_\text{irr} \rangle[\delta\Phi_\text{e}(t) - \delta\Phi_\text{e}(0)]\,.
\end{align}
The corresponding transition probability, averaged over the noise realizations, yields
\begin{equation}\label{offsetterm}
   \left\langle |\langle g^0(m_\text{l},m_\text{r})|\psi(t)\rangle|^2  \right\rangle_\text{av} = \eta^2|\langle e^0_\text{irr} | n | g^0_\text{irr} \rangle|^2 8\sigma^2e^2/\hbar^2\,,
\end{equation}
where we have approximated $\langle \delta \Phi_\text{e}(0) \delta \Phi_\text{e}(t) \rangle \approx 0$, valid for $t \gg t_\text{c}$~\cite{Clerk2010}. Equation \eqref{offsetterm} reproduces the unphysical offset term we obtained in Eq.~\eqref{corr} for $\langle C(0,t)\rangle_\text{av}$.

In summary, we find that fluctuations of the external flux lead to decay of the qubit with a gauge-invariant rate. In addition, the spurious offset term is also eliminated by enforcing the irrotational constraint. The latter removes the term $\propto\dot{\Phi}_\text{e}$ in the Lagrangian and Hamiltonian. A similar step to avoid  $\propto\dot{\Phi}_\text{e}$ terms was taken for a four-junction circuit by Qiu \textit{et al.}\ \cite{Qiu2016}. We conclude that Hamiltonians defined in terms of irrotational degrees of freedom represent the correct formulation for consistent calculations of multi-time correlation functions.

For the asymmetric dc SQUID, employing the irrotational variable specified by Eq.~\eqref{irrot} yields the Hamiltonian
\begin{align}\label{irrHSQUID}
    \mathcal{H}_\text{irr} =& 4E_\text{C}n^2 \\\nonumber
    &- E_\text{Jl} \cos{\left(\widetilde{\varphi}-\dfrac{C_\text{r}}{C_\Sigma}\varphi_\text{e}\right)} - E_\text{Jr} \cos{\left(-\widetilde{\varphi}-\dfrac{C_\text{l}}{C_\Sigma}\varphi_\text{e}\right)}\,. 
\end{align}
Note that the external flux is grouped with both junctions' potential energy terms, with weights given by capacitance ratios. 

\subsection{Circuit with inductor: fluxonium qubit}

\begin{figure}
\includegraphics[width=0.3\textwidth]{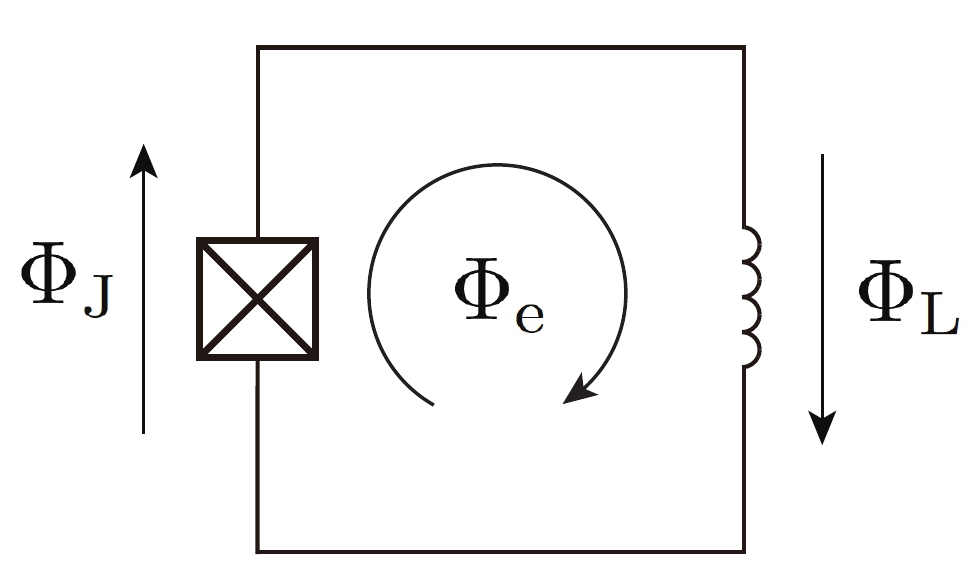}
\caption{\label{fig:fluxonium} The fluxonium qubit is formed by a Josephson junction and a superinductor with corresponding branch variables $\Phi_\text{J}$ and $\Phi_\text{L}$. An external flux $\Phi_\text{e}$ threads the loop.}
\end{figure}

The SQUID circuit analyzed in Sec.~\ref{sec-DC-SQUID} is an example of a single-loop circuit composed of Josephson junctions only. To see how flux distributes among Hamiltonian terms if the loop contains an inductor $L$, we consider the fluxonium qubit \cite{Manucharyan2009} (Fig.~\ref{fig:fluxonium}). We can obtain the irrotational degree of freedom by associating an auxiliary parallel capacitance $C_\text{L}$ with the inductor, and subsequently taking the limit $C_\text{L}\to0$.

The analysis of the circuit then proceeds in a manner analogous to that of the SQUID circuit. Following the same steps as above, we arrive at the Hamiltonian
\begin{equation*}
	\mathcal{H}_\text{aux} = 4 E_\text{C} n^2 - E_\text{J} \cos{\left(\widetilde{\varphi}-\dfrac{C_\text{L}}{C_\Sigma}\varphi_\text{e}\right)} 
    + \dfrac{1}{2}E_\text{L} \left(-\widetilde{\varphi}-\dfrac{C_\text{J}}{C_\Sigma}\varphi_\text{e}\right)^2 
\end{equation*}
with  $E_\text{L}=(\Phi_0/2\pi)^2/L$. Taking the limit $C_\text{L}\to0$ yields the proper irrotational Hamiltonian, \begin{equation}
	\mathcal{H}_\text{irr} = 4 E_{\text{CJ}}n^2
	- E_\text{J}\cos{\widetilde{\varphi}} 
	+ \dfrac{1}{2}E_\text{L}(-\widetilde{\varphi}-\varphi_\text{e})^2\,,
\end{equation}
where $E_{\text{CJ}}=e^2/2C_\text{J}$.
Thus, in a single-loop circuit with an inductor, the flux is entirely associated with the potential energy of the inductor.

\section{Canonical quantization of a general circuit network with external flux $\Phi_\text{e}(t)$}\label{sec:general}
In this section, we generalize the results obtained in the previous section to cover arbitrary circuits threaded by time-dependent flux. We start with single-loop circuits, and then extend to multi-loop circuits.

\subsection{Single-loop circuits\label{singleloop}}
For a general single-loop circuit with $N$ elements, we define the vector $\bm{\Phi} = (\Phi_1,\Phi_2,\ldots,\Phi_N)^\intercal$ collecting all branch fluxes in the circuit. Due to fluxoid quantization there are $N-1$ degrees of freedom denoted by $\widetilde{\bm{\Phi}}=(\widetilde{\Phi}_1,\widetilde{\Phi}_2,\ldots,\widetilde{\Phi}_{N-1})^\intercal$. In general, they are linear combinations of the original branch variables,
\begin{equation}
	\widetilde{\bm{\Phi}} = \mathbf{M} \bm{\Phi}\,,
\end{equation}
where $\mathbf{M}$ is an $(N{-}1){\times}N$ matrix with elements yet to be determined. It is useful to cast the fluxoid quantization constraint into matrix form as well,
\begin{equation}\label{fluxconstraint}
	\Phi_\text{e} = \mathbf{R} \bm{\Phi}\,,
\end{equation}
where $\mathbf{R}=(1,1,\dots,1)$ is a $1{\times} N$ matrix. We further introduce the augmented vector $\widetilde{\bm{\Phi}}_+$ and the augmented $N{\times} N$ matrix $\mathbf{M}_+$ by defining
\begin{equation}
    \bm{\widetilde{\Phi}}_+
    =
    \begin{pmatrix}
        \widetilde{\bm{\Phi}} \\
        \Phi_\text{e}
    \end{pmatrix}\,,
    \qquad
    \mathbf{M}_+
    =
    \begin{pmatrix}
    	\mathbf{M}\\
    	\mathbf{R}
    \end{pmatrix}\,.
\end{equation}
Both the information on degrees of freedom and on the fluxoid quantization constraint are now compactly written as
\begin{equation}
	\widetilde{\bm{\Phi}}_+ = \mathbf{M}_+ \bm{\Phi}\,.
\end{equation}
We note that $\det(\mathbf{M}_+){\neq} 0$ which is seen as follows. All rows of $\mathbf{M}$ generate genuine degrees of freedom which by definition must be linearly independent. Moreover, rows of $\mathbf{M}$ must be linearly independent of the row vector $\mathbf{R}$, since $\mathbf{R}$ does not generate a dynamical variable. Hence all rows of $\mathbf{M}_+$ are linearly independent.

We first consider a circuit loop composed of junctions only, and include capacitors and inductors subsequently. To construct irrotational degrees of freedom and find $\mathcal{H}_\text{irr}$, we inspect the kinetic energy,
\begin{equation}
\mathcal{L}_\text{k} = \dfrac{1}{2}\dot{\bm{\Phi}}^\intercal \mathbf{C} \dot{\bm{\Phi}}\,,
\end{equation}
where $\mathbf{C}=\text{diag}(C_\text{1},\,C_\text{2},\ldots,\,C_\text{N})$ is a diagonal matrix composed of all  capacitances in the circuit.  Since $\mathbf{M}_+$ is non-singular, we have 
\begin{equation}\label{phiphitilde}
\bm{\Phi}  = \mathbf{M}_+^{-1} \widetilde{\bm{\Phi}}_+\,,
\end{equation}
in which case the kinetic energy can be rewritten as 
\begin{equation}
	\mathcal{L}_\text{k} = \dfrac{1}{2}\dot{\bm{\widetilde{\Phi}}}_+^\intercal \mathbf{C}_\text{eff} \dot{\bm{\widetilde{\Phi}}}_+
\,,
\end{equation}
where the effective capacitance matrix is given by $\mathbf{C}_\text{eff}=(\mathbf{M}_+^{-1})^\intercal \mathbf{C} \mathbf{M}_+^{-1}$. For a loop composed of junctions both $\mathbf{C}$ and $\mathbf{C}_\text{eff}$ are invertible.

Following the discussion in Sec.~\ref{sec:time_dependent}, we obtain irrotational degrees of freedom by demanding that all Lagrangian terms $\propto\dot{\Phi}_\text{e}$ vanish. This is equivalent to requiring $\mathbf{C}_\text{eff}$ and, hence, its inverse to be block-diagonal. Specifically, by block matrix multiplication one finds
\begin{equation*}
\mathbf{C}_\text{eff}^{-1} = 
\begin{pmatrix}
	\mathbf{M}\\
	\mathbf{R}
\end{pmatrix}\!
\mathbf{C}^{-1}\!
\begin{pmatrix}
	\mathbf{M}^\intercal & \mathbf{R}^\intercal
\end{pmatrix}
= 
\begin{pmatrix}
	\mathbf{M}\mathbf{C}^{-1}\mathbf{M}^\intercal & \mathbf{M}\mathbf{C}^{-1}\mathbf{R}^\intercal \\
	\mathbf{R}\mathbf{C}^{-1}\mathbf{M}^\intercal & \mathbf{R}\mathbf{C}^{-1}\mathbf{R}^\intercal \\
\end{pmatrix},\vspace*{1mm}
\end{equation*}
so that the condition for irrotational degrees of freedom can be written as
\begin{equation}\label{m_eq}
	\mathbf{R}\mathbf{C}^{-1}\mathbf{M}^\intercal=(C_1^{-1},C_2^{-1},\ldots,C_N^{-1})\mathbf{M}^\intercal=0\,.
\end{equation}
This generalizes Eq.~\eqref{irrcondition}.
One readily verifies that a particular solution for $\mathbf{M}$ is given by 
\begin{equation}
    (\overline{\mathbf{M}})_{ij}= \delta_{ij} - \dfrac{C_i^{-1}}{\sum_{k=1}^N C_k^{-1}}\,.
\end{equation}
Generally speaking, solutions $\mathbf{M}^\intercal$ to Eq.~\eqref{m_eq} are matrices whose columns form a basis of the null space of $\mathbf{R}\mathbf{C}^{-1}$. Since the columns of $\overline{\mathbf{M}}^\intercal$ evidently form one particular basis of the null space, all possible solutions can be expressed as 
\begin{equation}
	\mathbf{M} = \mathbf{A} \overline{\mathbf{M}}\,.
\end{equation}
Here, $\mathbf{A}$ is an arbitrary non-singular $(N-1){\times}(N-1)$ matrix.

The Hamiltonian of the circuit is now obtained from the branch-flux Lagrangian by calculating the inverse $\mathbf{M}_+^{-1}$ and employing Eq.~\eqref{phiphitilde}. For $\mathbf{M}=\overline{\mathbf{M}}$, the inverse $\mathbf{M}_+^{-1}$ can be evaluated analytically, and Legendre transform of the Lagrangian yields
\begin{widetext}
    \begin{equation}\label{bigH}
        \mathcal{H}_\text{irr} =  \dfrac{1}{2}\sum_{i,j=1}^{N-1}\left[\dfrac{\delta_{ij}}{C_i}-\dfrac{1}{C_iC_j\sum_k C_k^{-1}}\right]Q_iQ_j 
         -\sum_{i=1}^{N-1} E_{\text{J}i} \cos{\left(\widetilde{\varphi}_i-\dfrac{C_i^{-1}}{\sum_k C_k^{-1}}\varphi_\text{e}\right)} 
         - E_{\text{J}N} \cos{\left( -\sum_{i=1}^{N-1}\widetilde{\varphi}_i -\dfrac{C_N^{-1}}{\sum_k C_k^{-1}}\varphi_\text{e} \right)}\,.
     \end{equation}
\end{widetext}
Hence, the weighting coefficient of the external flux for the $i$th element in the circuit loop is $C_i^{-1}/\sum_{k=1}^N C_k^{-1}$. (Note that the matrix $\mathbf{A}$ cannot affect the flux allocation.) It is simple to verify that the above Hamiltonian reduces to the Hamiltonian of the asymmetric SQUID Eq.~\eqref{irrHSQUID} when choosing $N=2$.

While our previous discussion only covers single-loop circuits consisting of Josephson junctions, it is not difficult to generalize to single-loop circuits including capacitors and inductors. If the loop includes capacitors, then $\mathbf{C}_\text{eff}$ remains invertible and we can proceed as before. As the only change, we must eliminate the corresponding junction potential terms from Eq.~\eqref{bigH}. (Details regarding the nature of the constraint replacing fluxoid quantization are provided in Appendix~\ref{sec:Kirchhoff}.)

In the presence of an inductor $L$ in a single-loop circuit, our results can be adapted by taking an appropriate limit as follows. Let the inductor be the $N$th element in the loop\footnote{Just as in Ref.~\cite{vool}, we focus on circuits with a simply-connected capacitive sub-network. This avoids the situation where a generalized velocity is absent from the Lagrangian. With this requirement, there can only be \emph{one} inductor in a single-loop circuit.}. We temporarily associate an auxiliary parallel capacitor $C_N$ with the inductor, perform time-dependent circuit quantization as described above, and then let the capacitance $C_N$ go to zero. This limit can be directly performed on the Hamiltonian \eqref{bigH}. Since the flux-grouping coefficients satisfy
\begin{equation}
    \dfrac{C_i^{-1}}{\sum_{k=1}^N C_k^{-1}} \xrightarrow{C_N\to0} \delta_{iN}\,,
\end{equation}
the external flux is entirely grouped with the inductor in the single-loop circuit, and the resulting $N$th potential energy term reads
\begin{equation}
    \dfrac{1}{2}E_\text{L}\left(-\sum_{i=1}^{N-1}\widetilde{\varphi}_i-\varphi_\text{e}\right)^2
\end{equation}
with $E_\text{L} = (\Phi_0/2\pi)^2/L$.

Strictly speaking, every circuit loop will have some finite geometric inductance $L$. Hence, one might wonder whether this inductance has to be included as it leads to different flux grouping. We show in Appendix~\ref{sec:geoinductance} that inclusion of negligible $L$ is not necessary and that the limit $L\to0$ is well-behaved. Specifically, the limit $L\to0$ leads back to the inductor-less case discussed above, with flux allocated to the various junction terms.

The above results are formulated for single-loop circuits, but actually apply more broadly. Often, only a single loop $\mathsf{L}$ inside a multi-loop circuit is threaded by a time-varying flux, while the flux through the remaining loops is zero. 
In cases where each element outside $\mathsf{L}$ is in parallel to an element inside $\mathsf{L}$, the multi-loop circuit effectively  reduces to a single-loop circuit.

\subsection{Multi-loop circuits} 
Next, we extend our discussion to the case of multi-loop circuits threaded by time-dependent fluxes through each loop. Some of our development here resembles the discussion by Burkard \textit{et al.}\ \cite{Burkard2004}. We begin by labeling all  circuit elements (i.e., Josephson junctions, capacitors, inductors) in the network, and construct a vector containing the $N$ branch flux variables associated with the elements, $\bm{\Phi} = (\Phi_1,\Phi_2,\ldots,\Phi_N)^\intercal$. Due to the constraints from fluxoid quantization and Faraday's law, these variables are in general not independent of each other. 

Constraints apply to loops in the circuit. To eliminate ambiguity in the choice of loops, we select all meshes, which are loops containing no other loops. Suppose the circuit encompasses $F$ meshes, then we define $\bm{\Phi}_{\text{e}}=(\Phi_{\text{e}}^1,\Phi_{\text{e}}^2,\ldots,\Phi_{\text{e}}^F)^\intercal$ as the vector containing all time-dependent external fluxes threading each mesh. The positive directions of branch fluxes and external fluxes is defined in the beginning in one consistent way maintained throughout the calculation. The number of constraints applying to the $N$ variables is $F$, leading to $N{-}F$ degrees of freedom in the circuit. 

\begin{figure}
\includegraphics[width=0.38\textwidth]{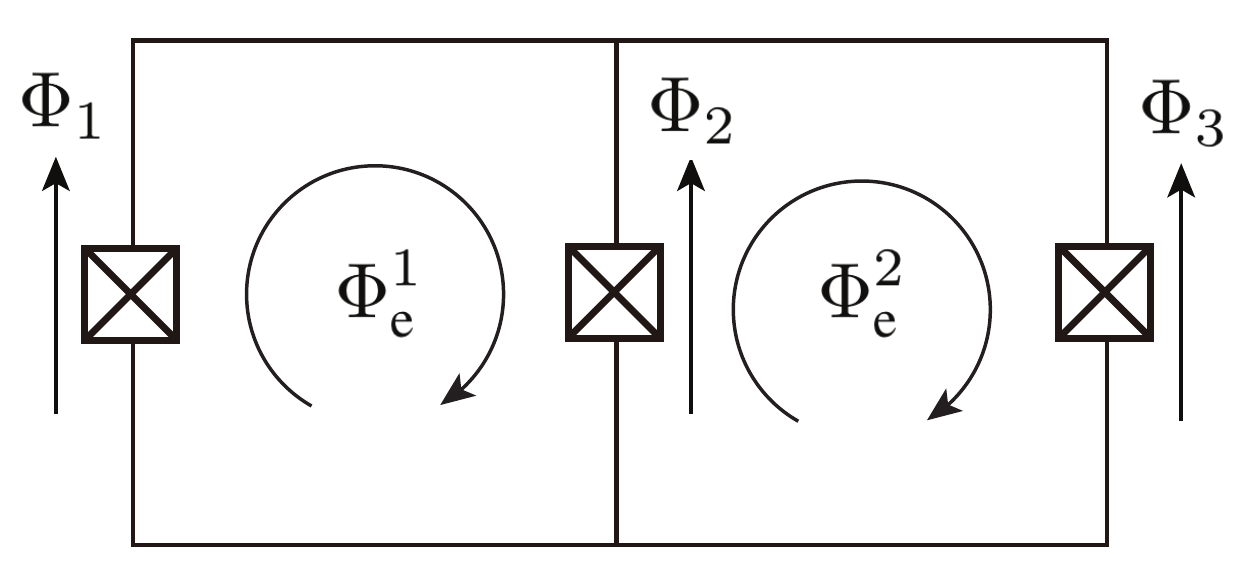}
\caption{\label{fig:dsquid} Multi-loop circuit formed by two SQUID loops with one shared junction. $\Phi_\text{1}$, $\Phi_\text{2}$, and $\Phi_\text{3}$ are branch flux variables for the three Josephson junctions. External fluxes $\Phi_\text{e}^1$ and $\Phi_\text{e}^2$ thread the left and right loops.}
\end{figure}

We denote the vector composed of the corresponding dynamical variables by $\bm{\widetilde{\Phi}}=(\widetilde{\Phi}_1,\widetilde{\Phi}_2,\ldots,\widetilde{\Phi}_{N-F})^\intercal$. 
These $N{-}F$ variables are given by appropriate linear combinations of the $N$ branch variables,
\begin{equation}
	\bm{\widetilde{\Phi}} = \mathbf{M} \bm{\Phi}\,.
\end{equation}
Here, $\mathbf{M}$ is an $(N{-}F){\times} N$ matrix with elements $(\mathbf{M})_{ij}$ to be determined. The constraints from fluxoid quantization and Faraday's law can jointly be expressed as 
\begin{equation}
	\bm{\Phi}_\text{e} = \mathbf{R} \bm{\Phi}\,,
\end{equation}
where the $F{\times} N$ mesh matrix $\mathbf{R}$ is defined as follows. Let $\mathsf{L}_i$ be the $i$th mesh threaded by flux $\Phi_\text{e}^i$. Then the elements of $\mathbf{R}$ are given by 
\begin{equation*}
    (\mathbf{R})_{ij}=
    \begin{cases}
      \phantom{-}1\,, & \text{$\Phi_j \in \mathsf{L}_i$ with same orientation as $\Phi_\text{e}^i$} \\
      -1\,, & \text{$\Phi_j \in \mathsf{L}_i$ with orientation opposite to $\Phi_\text{e}^i$} \\
      \phantom{-}0\,, & \text{$\Phi_j \notin \mathsf{L}_i$}
    \end{cases}\,,
  \end{equation*}
where the orientation of $\Phi_\text{e}^i$ is clockwise (counterclockwise) when the local magnetic field points into (out of) the plane of the loop (see, e.g.~the circular arrows in Fig.~{\ref{fig:dsquid}}). Analogous to the single-loop case, we further introduce the augmented vector $\bm{\widetilde{\Phi}}_+$ and the $N{\times} N$ matrix $\mathbf{M}_+$ via
\begin{equation}
    \bm{\widetilde{\Phi}}_+
    =
    \begin{pmatrix}
        \bm{\widetilde{\Phi}} \\
        \bm{\Phi}_\text{e}
    \end{pmatrix}\,,
    \qquad
    \mathbf{M}_+
    =
    \begin{pmatrix}
    	\mathbf{M}\\
    	\mathbf{R}
    \end{pmatrix}\,,
\end{equation}
so that both information on degrees of freedom and constraints is again compactly captured by
\begin{equation}
	\bm{\widetilde{\Phi}}_+ = \mathbf{M}_+ \bm{\Phi}\,.
\end{equation}
For the same reasons as in the single-loop case, we have $\det(\mathbf{M}_+){\neq} 0$.

To identify the irrotational degrees of freedom, we turn to the kinetic energy
\begin{equation}
	\mathcal{L}_\text{k} = \dfrac{1}{2}\dot{\bm{\widetilde{\Phi}}}_+^\intercal \mathbf{C}_\text{eff} \dot{\bm{\widetilde{\Phi}}}_+\,,
\end{equation}
written in terms of $\mathbf{C}_\text{eff} = (\mathbf{M}_+^{-1})^\intercal \mathbf{C} \mathbf{M}_+^{-1}$. Irrotational degrees of freedom are obtained by making the terms $\propto\dot{\Phi}_\text{e}$ in the Lagrangian vanish. As before, this leads to the condition
\begin{equation}
	\mathbf{R}\mathbf{C}^{-1}\mathbf{M}^\intercal=0\,.
\end{equation}

While $\mathbf{R}$ is very simple in the single-loop case, its structure is now more complex as it encodes the geometry of the multi-loop circuit.
Different from the single-loop problem, this added complexity makes it impractical to obtain general closed expressions for $\mathbf{M}$. Instead, a particular solution for $\mathbf{M}$ can be obtained on a case-by-case basis as follows. Perform the singular-value decomposition
\begin{equation}
    \mathbf{R}\mathbf{C}^{-1} = \mathbf{U} \mathbf{S} \mathbf{V}\,,
\end{equation}
where $\mathbf{V}$ is an $N{\times} N$ matrix. $\overline{\mathbf{M}}$ can be obtained from the transpose of the last $(N{-}F)$ columns of $\mathbf{V}$, and the general solution satisfies $\mathbf{M} = \mathbf{A} \overline{\mathbf{M}}$ with $\mathbf{A}$ an arbitrary non-singular $(N{-}F){\times}(N{-}F)$ matrix. 

As before, Lagrangian and Hamiltonian are constructed by calculating $\mathbf{M}_+^{-1}$ and expressing branch variables in terms of the irrotational degrees of freedom. The grouping of flux $\Phi_\text{e}^j$ with the $i$th potential energy term can be obtained from $(\mathbf{M}_+^{-1})_{i,N-F+j}$. For capacitors and inductors in the multi-loop circuit we follow the same procedures discussed in the single-loop analysis. The above formulation likewise applies to multi-loop circuits threaded by both time-dependent and time-independent flux,  considering static flux a special case of time-varying flux.

For illustration, we apply the above formalism to a concrete multi-loop circuit: two SQUID loops with a shared junction (Fig.~\ref{fig:dsquid}). The junction capacitances and Josephson energies are denoted $C_k$ and $E_{\text{J}k}$ $(k=1,2,3)$. The left and right loops are threaded by external flux $\Phi_\text{e}^1$ and $\Phi_\text{e}^2$, respectively. The capacitance matrix and the mesh matrix for the given circuit-element orientations are 
\begin{equation}
    \mathbf{C}
    = \text{diag}(C_1,C_2,C_3)\,, \qquad
    \mathbf{R}
    = 
    \begin{pmatrix}
        1 & -1 & 0 \\
        0 & 1 & -1 \\
    \end{pmatrix}\,.
\end{equation}
Singular-value decomposition and re-scaling yields a particular solution of Eq.~\eqref{multi_cond}, namely
\begin{equation}
    \mathbf{M} = C_\Sigma^{-1}
    \begin{pmatrix}
        C_1 & C_2 & C_3
    \end{pmatrix}\,,
\end{equation}
where $C_\Sigma=\sum_{k=1}^3 C_k$. With this, one obtains
\begin{equation}
    \mathbf{M}_+^{-1}=
    \begin{pmatrix}
        1 & (C_2+C_3)C_\Sigma^{-1} & C_3C_\Sigma^{-1} \\
        1 & -C_1C_\Sigma^{-1} & C_3C_\Sigma^{-1} \\
        1 & -C_1C_\Sigma^{-1} & -(C_1+C_2)C_\Sigma^{-1} \\
    \end{pmatrix}\,,
\end{equation}
and 
\begin{equation}\label{multi_cond}
    \mathbf{C}_\text{eff} = 
    \begin{pmatrix}
    C_\Sigma & 0 & 0 \\
    0 & C_1(C_2+C_3)C_\Sigma^{-1} & C_1C_3C_\Sigma^{-1} \\
    0 & C_1C_3C_\Sigma^{-1} & (C_1+C_2)C_3C_\Sigma^{-1} \\
    \end{pmatrix}\,.
\end{equation}
This now allows us to construct the irrotational Hamiltonian
\begin{align}
    \mathcal{H}_\text{irr} = &  4E_\text{C}n^2 \nonumber \\
    & - E_\text{J1} \cos{\left(\widetilde{\varphi}+\dfrac{C_2+C_3}{C_\Sigma}\varphi_\text{e}^1+\dfrac{C_3}{C_\Sigma}\varphi_\text{e}^2\right)} \nonumber \\
    & - E_\text{J2} \cos{\left(\widetilde{\varphi}-\dfrac{C_1}{C_\Sigma}\varphi_\text{e}^1+\dfrac{C_3}{C_\Sigma}\varphi_\text{e}^2\right)} \nonumber \\
    & - E_\text{J3} \cos{\left(\widetilde{\varphi}-\dfrac{C_1}{C_\Sigma}\varphi_\text{e}^1-\dfrac{C_1+C_2}{C_\Sigma}\varphi_\text{e}^2\right)}\,,
\end{align}
where $E_\text{C} = e^2/2C_\Sigma$.

\section{Conclusions}\label{sec:summary}
Standard circuit quantization \cite{qnetwork,devoret1995quantum, Burkard2004,vool,Ulrich2016} applies to circuits threaded by static external flux, but leads to inconsistencies when applied to circuits subject to time-dependent flux.
Here, we have presented a generalization of circuit quantization for single-loop and multi-loop circuits that incorporates time-dependent flux. Our treatment shows that time-varying flux generally produces Lagrangian and Hamiltonian terms $\propto\dot{\Phi}_\text{e}$ which are crucial for resolving inconsistencies. We find that the freedom in choosing independent circuit variables leaves expectation values of single-time observables invariant, but generates gauge-dependent offsets in multi-time correlation functions. 
We have identified irrotational degrees of freedom which both eliminate the terms $\propto\dot{\Phi}_\text{e}$ in the Hamiltonian and remove the spurious offset terms in correlation functions associated with relaxation and pure-dephasing dynamics. Finally, we note that the irrotational constraint enforces a unique allocation of external flux to the various potential energy terms in the Hamiltonian.

\begin{acknowledgments}
We are indebted to Peter Groszkowski, Andy C.~Y.~Li, Yao Lu, and Jay Lawrence for illuminating discussions. This work was supported by the Northwestern--Fermilab Center for Applied Physics and Superconducting Technologies. 
The work of J.~A.~Sauls was supported by National Science Foundation Grant DMR-1508730.
\end{acknowledgments}

\appendix

\section{Evaluation of  $\langle C(0,t)\rangle_\text{av}$ \label{sec:correlation}} 
This appendix details the evaluation of the noise-averaged transition probability
introduced in Sec.~\ref{sec:correlator},
\begin{equation}\label{cavg1}
\langle C(0,t) \rangle_\text{av} = \Big\langle |\langle \psi(0)|\psi(t)\rangle|^2 \Big\rangle_\text{av}\,.
\end{equation}
This quantity is defined for a particular choice of the SQUID's degree of freedom, specified by $(m_\text{l},m_\text{r})$. As in the main text, the initial state is the unperturbed excited state $|e^0(m_\text{l},m_\text{r})\rangle$, an eigenstate of $\mathcal{H}^0(m_l,m_r)$. (In the following, we will simplify notation and frequently suppress the dependence on $m_\text{l}$, $m_\text{r}$ where context allows.)

Truncating the Hilbert space to the subspace spanned by ground and first excited state, we can rewrite 
\begin{equation}
C(0,t) = |\langle \psi(0) | \psi(t) \rangle|^2 = 1 - |\langle g^0 | \psi(t) \rangle|^2\,,
\end{equation}
and calculate the overlap with the ground state perturbatively. This yields
\begin{equation}
\langle g^0 | \psi(t) \rangle = -\dfrac{i}{\hbar}\int_0^t e^{-i\omega_\text{eg}t'} V(t') dt'\,,
\end{equation}
where the perturbation term in the integrand is
\begin{align}
V(t) = &  \langle g^0 | \partial_{\Phi_\text{e}} \mathcal{H}(m_\text{l},m_\text{r})| e^0\rangle \delta\Phi_\text{e}(t)  \nonumber\\
& + \bar{\eta} \langle g^0 | Q |e^0\rangle \delta\dot{\Phi}_\text{e} (t) \nonumber\\
= &  \underbrace{\langle g^0_\text{irr} | \partial_{\Phi_\text{e}} \mathcal{H}_\text{irr}| e^0_\text{irr}\rangle \delta\Phi_\text{e}(t)}_{V_\text{irr}(t)} \nonumber\\
& + \underbrace{\eta\langle g^0_\text{irr} | Q | e^0_\text{irr} \rangle [- i\omega_\text{eg}\delta\Phi_\text{e}(t) + \delta\dot{\Phi}_\text{e} (t)]}_{W(t)}\,.
\end{align}
The last expression follows from applying Eq.~\eqref{Utransform}. We note immediately that $V_\text{irr}(t)$ is precisely the perturbation term obtained within the irrotational frame of reference. Plugging back into Eq.~\eqref{cavg1}, we obtain
\begin{widetext}
	\begin{equation}
	\langle C(0,t) \rangle_\text{av} = 1-\dfrac{1}{\hbar^2}\int_0^{t}\!\!\!dt_1\int_0^{t}\!\!\!dt_2\,e^{-i\omega_\text{eg}(t_1-t_2)}\left[ 
	\langle V_\text{irr}(t_1)V^*_\text{irr}(t_2) \rangle_\text{av} + \langle V_\text{irr}(t_1)W^*(t_2) + W(t_1) V_\text{irr}^*(t_2) \rangle_\text{av}
	+ \langle W(t_1)W^*(t_2) \rangle_\text{av}
	\right]\,.
	\end{equation}
	Here, the term proportional to $\delta\Phi_\text{e}(t)$ in $W$ is canceled upon performing integration by parts on the $\delta\dot{\Phi}_\text{e}(t)$ term.
	For classical noise with zero mean we have $\langle \delta\Phi_\text{e}(t)\rangle_\text{av}=0$, and the noise spectrum is symmetric, $S_{\Phi_\text{e}}(\omega)=S_{\Phi_\text{e}}(-\omega)$. As a consequence, one finds that the terms mixing $V_\text{irr}$ and $W$ cancel out. We evaluate the remaining two terms separately. For the $V_\text{irr}$ correlator we find
	\begin{align}
	\dfrac{1}{\hbar^2}\int_0^{t}\!\!\!dt_1\int_0^{t}\!\!\!dt_2\,e^{-i\omega_\text{eg}(t_1-t_2)} 
	\langle V_\text{irr}(t_1)V^*_\text{irr}(t_2) \rangle_\text{av}
	&=\dfrac{1}{\hbar^2} |\langle g^0_\text{irr} | \partial_{\Phi_\text{e}} \mathcal{H}_\text{irr} | e^0_\text{irr}\rangle|^2 \int_0^{t}\!\!\!dt_1\int_0^{t}\!\!\!dt_2\, e^{-i\omega_\text{eg}(t_2-t_1)}\langle \delta\Phi_\text{e}(t_2-t_1)\delta\Phi_\text{e}(0)\rangle \nonumber\\
	&\approx \dfrac{1}{\hbar^2}|\langle g^0_\text{irr} | \partial_{\Phi_\text{e}} \mathcal{H}_\text{irr} | e^0_\text{irr}\rangle|^2 S_{\Phi_\text{e}}(\omega_{\text{eg}})\,t\,,
	\end{align}
	where we have used  $t \gg t_\text{c}$ (correlation time of the noise) for the approximation in the last step \cite{Clerk2010}. The $W$ correlator yields
	\begin{align}
	\dfrac{1}{\hbar^2}\int_0^{t}\!\!\!dt_1\int_0^{t}\!\!\!dt_2\,e^{-i\omega_\text{eg}(t_1-t_2)} 
	\langle W_\text{irr}(t_1)W^*_\text{irr}(t_2) \rangle_\text{av} = & \dfrac{1}{\hbar^2}\eta^2|\langle g^0_\text{irr} | Q | e^0_\text{irr} \rangle|^2
	\langle [\delta\Phi_\text{e}(t) e^{-i\omega t}-\delta\Phi_\text{e}(0)]\times[\text{c.c.}] \rangle_\text{av}\nonumber\\
	\approx & \dfrac{1}{\hbar^2}\eta^2|\langle g^0_\text{irr} | Q | e^0_\text{irr} \rangle|^2 2\sigma^2\,,
	\end{align}
\end{widetext}
using the same type of approximation as above. Together, these results confirm the validity of Eq.~\eqref{corr} for $t_\text{c}\ll t\ll T_1$.

\section{Constraint in single-loop circuits with capacitors\label{sec:Kirchhoff}}
Fluxoid quantization applies to closed loops, consisting of Josephson junctions and inductors. In general, a loop may be interrupted by a capacitor. In this case, fluxoid quantization is not applicable, but Faraday's law leads to a similar constraint for time-dependent flux \cite{Burkard2004}. To see this, consider the example of an LC circuit threaded by an external flux $\Phi_\text{e}$. If this flux is time-dependent, Faraday's law results in the relation
\begin{equation}
V_\text{C}+V_\text{L} = -\dot{\Phi}_\text{e}(t)\,,
\end{equation}
where $V_\text{C}$ and $V_\text{L}$ are the voltages across the capacitor and inductor, respectively.
The time integral of this equation leads to the following constraint on the branch flux variables:
\begin{equation}
\Phi_\text{C} + \Phi_\text{L} = -\Phi_\text{e}(t)\,.
\end{equation}
Once we redefine $\Phi_\text{e}$ to absorb the sign from Lenz's rule, this constraint has the same form as the fluxoid-quantization constraint, see Eq.~\eqref{fluxconstraint}. It is straightforward to extend this result from the LC circuit to a general mesh in a circuit network. Thus, open loops involving capacitors can be treated on equal footing with the closed loops subject to fluxoid quantization.

\section{Time of pure dephasing due to flux noise}
\label{sec:pure_dephasing_time}
In the main text, we focused on flux-noise--induced relaxation of the qubit, and illustrated that inconsistencies may arise if flux is not grouped as prescribed by the irrotational frame of reference. We here briefly note that calculations of pure-dephasing times are likewise affected by the choice of the degree of freedom. For simplicity, we again ignore the $1/f$ nature of flux noise. Working in the irrotational frame, we indeed find the standard expression \cite{Clerk2010,Groszkowski2017b} for the decay of the off-diagonal density matrix elements,
\begin{equation}
\langle \rho_\text{eg} (t) \rangle \approx \exp[-(\partial_{\Phi_\text{e}}\omega_\text{eg})^2 S_{\Phi_\text{e}}(0)\,t/2 ]\,.
\end{equation}
By contrast, when choosing other degrees of freedom specified by $(m_\text{l},m_\text{r})$, an additional frame-dependent term arises, 
\begin{align}
\langle \rho'_\text{eg} (t) \rangle \approx & \exp[-(\partial_{\Phi_\text{e}}\omega_\text{eg})^2 S_{\Phi_\text{e}}(0)\,t/2 ] \\\nonumber
& \times \exp[-(\langle g | Q | g \rangle - \langle e | Q | e \rangle)^2\eta^2(m_\text{l},m_\text{r})\sigma^2/\hbar^2]\,.
\end{align}
As for the relaxation-time results, the spurious offset is eliminated by choosing the irrotational frame where the Lagrangian term $\propto\dot{\Phi}_\text{e}$ vanishes and $\eta=0$.

\section{Geometric inductance and $L\to0$ limit}
\label{sec:geoinductance}
Section \ref{sec:general} showed that flux through loops composed of Josephson junctions is distributed across the various junction terms. By contrast, for loops including an inductor the flux is grouped with the inductor term. This situation raises two questions: Does the finite geometric inductance associated with any loop have to be included in circuit quantization even if it is negligibly small, and is the limit $L\to0$  singular? In this appendix we prove that both questions can be answered in the negative.  

Consider a circuit loop composed of geometric inductance $L$, and $N$ Josephson junctions. Denote by $\Phi_\text{L}$ the branch flux across the inductor, and by $\Phi_1,\ldots,\Phi_N$ the junction branch variables. Fluxoid quantization leaves us with $N$ degrees of freedom. We may associate one of them with the inductor variable, $\widetilde{\Phi}_N=-\Phi_\text{L}+\Phi_\text{e}$, grouping the external flux exclusively with the inductor. The remaining $N{-}1$ degrees of freedom are composed of linear combinations of $\Phi_1,\ldots,\Phi_N$.

Generally, the resulting Lagrangian will have kinetic terms $\propto$ $\dot{\widetilde{\Phi}}_i\dot{\widetilde{\Phi}}_N$ ($i{\neq}N$) which couple the inductor to the remaining degrees of freedom. To evaluate the $L{\to}0$ limit, it is crucial to eliminate this coupling \cite{Qiu2016}. We can do so in a way analogous to Sec.~\ref{singleloop}. The fluxoid quantization condition is here replaced by $\sum_{i=1}^N \Phi_i=\widetilde{\Phi}_N$, i.e., the role played by $\Phi_\text{e}$ in  Eq.~\eqref{bigH} is now played by $\widetilde{\Phi}_N$. Defining shorthands $\bm{\widetilde{\Phi}}=(\widetilde{\Phi}_1,\ldots,\widetilde{\Phi}_{N-1})$ and $\mathbf{Q}=(Q_1,\ldots,Q_{N-1})$, we obtain for the Hamiltonian 
\begin{equation}
\mathcal{H} = \mathcal{H}_\text{irr}(\mathbf{Q},\bm{\widetilde{\Phi}}, \widetilde{\Phi}_N) + \mathcal{H}_\text{osc}\,,
\end{equation}
where $\mathcal{H}_\text{irr}(\mathbf{Q},\bm{\widetilde{\Phi}}, \widetilde{\Phi}_N)=[\text{Eq.}\ \eqref{bigH}]|_{\Phi_\text{e}=\widetilde{\Phi}_N}$.
The oscillator Hamiltonian composed of inductor and series combination of all junction capacitors is
\begin{equation}
\mathcal{H}_\text{osc} = 4E_\text{C}n_N^2 + \dfrac{1}{2}E_\text{L}(\widetilde{\varphi}_N-\varphi_\text{e})^2 - E_0
\end{equation}
with $E_\text{C} = \dfrac{1}{2}e^2\sum_k C_k^{-1}$ and $E_\text{L} = (\Phi_0/2\pi)^2/L$. To prevent divergence of the oscillator ground state energy in the $L\to0$ limit, we explicitly subtract its value $E_0=\dfrac{1}{2}\sqrt{8E_\text{L}E_\text{C}}$.

In the limit $L\to0$ where $E_\text{L}\to\infty$, the oscillator will occupy its ground state. Moreover, position fluctuations around $\widetilde{\varphi}_N=\varphi_\text{e}$ will become negligible. As a result, we find the limiting behavior
\begin{equation}
\mathcal{P}_0\mathcal{H}\mathcal{P}_0 
\xrightarrow{E_\text{L}\to \infty}
\mathcal{H}_\text{irr}(\mathbf{Q},\bm{\widetilde{\Phi}},\Phi_\text{e})\,,
\end{equation}
where $\mathcal{P}_0$ projects onto the oscillator ground state. We thus find that the $L\to0$ limit reproduces the irrotational Hamiltonian we obtained in Sec.~\ref{singleloop} without the inductor. In other words, the $L\to0$ limit is non-singular, and sufficiently small loop inductances may be neglected in the circuit quantization. 

As a final check we consider the decay rate $\Gamma_1 \propto |\langle g| \partial_{\Phi_\text{e}} \mathcal{H}   | e\rangle|^2$. It is tempting to reason that for a circuit loop with an inductor (where flux is grouped with the inductor term), one simply finds $\Gamma_1\propto E_\text{L}^2$ (wrong). To see that the $E_\text{L}\to\infty$ limit is well-behaved, we consider flux noise as denoted in Eq.~\eqref{noise}. Taking the flux derivative, we see that the decay rate is proportional to $E_\text{L}^2 |\langle g^0 |  \widetilde{\varphi}_N | e^0 \rangle|^2$, where $|g^0\rangle$ and $|e^0\rangle$ are eigenstates of $\mathcal{H}$ for $\Phi_\text{e} = \Phi_\text{e}^0$. Despite this form, $\Gamma_1$ does not diverge for $E_\text{L}\to \infty$, since both eigenstates and the operator $\widetilde{\varphi}_N$ depend on $E_\text{L}$ in such a way that $E_\text{L}$ factors cancel. We show in the following that the decay rate reaches a proper limit as the geometric inductance approaches zero.

We saw above that fluctuations of $\widetilde{\varphi}_N$ around $\varphi_\text{e}$ become negligible in the $E_\text{L}\to0$ limit. Hence, $\widetilde{\varphi}_N - \varphi_\text{e}$ is a small parameter. We rewrite $\widetilde{\varphi}_N$ as
$\widetilde{\varphi}_N = \varphi_\text{e} + (\widetilde{\varphi}_N - \varphi_\text{e})$
and plug this into the Hamiltonian to obtain
\begin{equation}
\mathcal{H} \approx \mathcal{H}_\text{irr}(\mathbf{Q},\bm{\widetilde{\Phi}},\Phi_\text{e}) + \mathcal{H}_\text{osc}
+ V\,,
\end{equation}
with the perturbation $V=(\partial_{\varphi_\text{e}}\mathcal{H}_\text{irr})(\widetilde{\varphi}_N - \varphi_\text{e})$. Employing perturbation theory, the eigenstates $|g^0\rangle$ and $|e^0\rangle$ can now be expanded to first order in $V$. One finds that the resulting corrections are inversely proportional to $\hbar\omega_\text{osc}=\sqrt{8E_\text{C}E_\text{L}}$ for large $E_\text{L}$. Dependence on $E_\text{L}$ is also hiding in  $\widetilde{\varphi}_N$ when expressed in terms of oscillator ladder operators,
$\widetilde{\varphi}_N \propto (a+a^\dag)E_\text{L}^{-1/4}$.
With these $E_\text{L}$ dependencies, the decay rate is seen to have a finite value in the limit $E_\text{L}\to\infty$. Furthermore, one finds
\begin{equation}
\langle g^0|E_\text{L}\widetilde{\varphi}_N|e^0\rangle|^2 \xrightarrow{E_\text{L}\to \infty} |\langle g_\text{irr}^0|\partial_{\varphi_\text{e}}\mathcal{H}_\text{irr}|e_\text{irr}^0\rangle|^2\,,
\end{equation}
where $|g_\text{irr}^0\rangle$ and $|e_\text{irr}^0\rangle$ are the eigenstates of the irrotational Hamiltonian $\mathcal{H}_\text{irr}(\mathbf{Q},\bm{\widetilde{\Phi}},\Phi_\text{e}^0)$. The RHS of the above equation reproduces the result obtained by neglecting the geometric inductance in the first place.

\bibliographystyle{apsrev4-1}
\bibliography{Manuscript.bib}

\begin{thebibliography}{32}%
\makeatletter
\providecommand \@ifxundefined [1]{%
 \@ifx{#1\undefined}
}%
\providecommand \@ifnum [1]{%
 \ifnum #1\expandafter \@firstoftwo
 \else \expandafter \@secondoftwo
 \fi
}%
\providecommand \@ifx [1]{%
 \ifx #1\expandafter \@firstoftwo
 \else \expandafter \@secondoftwo
 \fi
}%
\providecommand \natexlab [1]{#1}%
\providecommand \enquote  [1]{``#1''}%
\providecommand \bibnamefont  [1]{#1}%
\providecommand \bibfnamefont [1]{#1}%
\providecommand \citenamefont [1]{#1}%
\providecommand \href@noop [0]{\@secondoftwo}%
\providecommand \href [0]{\begingroup \@sanitize@url \@href}%
\providecommand \@href[1]{\@@startlink{#1}\@@href}%
\providecommand \@@href[1]{\endgroup#1\@@endlink}%
\providecommand \@sanitize@url [0]{\catcode `\\12\catcode `\$12\catcode
  `\&12\catcode `\#12\catcode `\^12\catcode `\_12\catcode `\%12\relax}%
\providecommand \@@startlink[1]{}%
\providecommand \@@endlink[0]{}%
\providecommand \url  [0]{\begingroup\@sanitize@url \@url }%
\providecommand \@url [1]{\endgroup\@href {#1}{\urlprefix }}%
\providecommand \urlprefix  [0]{URL }%
\providecommand \Eprint [0]{\href }%
\providecommand \doibase [0]{http://dx.doi.org/}%
\providecommand \selectlanguage [0]{\@gobble}%
\providecommand \bibinfo  [0]{\@secondoftwo}%
\providecommand \bibfield  [0]{\@secondoftwo}%
\providecommand \translation [1]{[#1]}%
\providecommand \BibitemOpen [0]{}%
\providecommand \bibitemStop [0]{}%
\providecommand \bibitemNoStop [0]{.\EOS\space}%
\providecommand \EOS [0]{\spacefactor3000\relax}%
\providecommand \BibitemShut  [1]{\csname bibitem#1\endcsname}%
\let\auto@bib@innerbib\@empty
\bibitem [{\citenamefont {Devoret}\ and\ \citenamefont
  {Schoelkopf}(2013)}]{Devoret2013}%
  \BibitemOpen
  \bibfield  {author} {\bibinfo {author} {\bibfnamefont {M.~H.}\ \bibnamefont
  {Devoret}}\ and\ \bibinfo {author} {\bibfnamefont {R.~J.}\ \bibnamefont
  {Schoelkopf}},\ }\href {\doibase 10.1126/science.1231930} {\bibfield
  {journal} {\bibinfo  {journal} {Science}\ }\textbf {\bibinfo {volume}
  {339}},\ \bibinfo {pages} {1169} (\bibinfo {year} {2013})}\BibitemShut
  {NoStop}%
\bibitem [{\citenamefont {Nguyen}\ \emph {et~al.}(2018)\citenamefont {Nguyen},
  \citenamefont {Lin}, \citenamefont {Somoroff}, \citenamefont {Mencia},
  \citenamefont {Grabon},\ and\ \citenamefont {Manucharyan}}]{Nguyen2018}%
  \BibitemOpen
  \bibfield  {author} {\bibinfo {author} {\bibfnamefont {L.~B.}\ \bibnamefont
  {Nguyen}}, \bibinfo {author} {\bibfnamefont {Y.-H.}\ \bibnamefont {Lin}},
  \bibinfo {author} {\bibfnamefont {A.}~\bibnamefont {Somoroff}}, \bibinfo
  {author} {\bibfnamefont {R.}~\bibnamefont {Mencia}}, \bibinfo {author}
  {\bibfnamefont {N.}~\bibnamefont {Grabon}}, \ and\ \bibinfo {author}
  {\bibfnamefont {V.~E.}\ \bibnamefont {Manucharyan}},\ }\href
  {http://arxiv.org/abs/1810.11006} {\bibfield  {journal} {\bibinfo  {journal}
  {arXiv:1810.11006}\ } (\bibinfo {year} {2018})}\BibitemShut {NoStop}%
\bibitem [{\citenamefont {Yurke}\ and\ \citenamefont
  {Denker}(1984)}]{qnetwork}%
  \BibitemOpen
  \bibfield  {author} {\bibinfo {author} {\bibfnamefont {B.}~\bibnamefont
  {Yurke}}\ and\ \bibinfo {author} {\bibfnamefont {J.~S.}\ \bibnamefont
  {Denker}},\ }\href {\doibase 10.1103/PhysRevA.29.1419} {\bibfield  {journal}
  {\bibinfo  {journal} {Phys. Rev. A}\ }\textbf {\bibinfo {volume} {29}},\
  \bibinfo {pages} {1419} (\bibinfo {year} {1984})}\BibitemShut {NoStop}%
\bibitem [{\citenamefont {Devoret}(1997)}]{devoret1995quantum}%
  \BibitemOpen
  \bibfield  {author} {\bibinfo {author} {\bibfnamefont {M.~H.}\ \bibnamefont
  {Devoret}},\ }\href@noop {} {\emph {\bibinfo {title} {Quantum Fluctuations,
  Les Houches, Session LXIII}}},\ edited by\ \bibinfo {editor} {\bibfnamefont
  {S.}~\bibnamefont {Reynaud}}, \bibinfo {editor} {\bibfnamefont
  {E.}~\bibnamefont {Giacobino}}, \ and\ \bibinfo {editor} {\bibfnamefont
  {J.}~\bibnamefont {Zinn-Justin}}\ (\bibinfo  {publisher} {Elsevier Science},\
  \bibinfo {year} {1997})\ pp.\ \bibinfo {pages} {351--386}\BibitemShut
  {NoStop}%
\bibitem [{\citenamefont {Burkard}\ \emph {et~al.}(2004)\citenamefont
  {Burkard}, \citenamefont {Koch},\ and\ \citenamefont
  {DiVincenzo}}]{Burkard2004}%
  \BibitemOpen
  \bibfield  {author} {\bibinfo {author} {\bibfnamefont {G.}~\bibnamefont
  {Burkard}}, \bibinfo {author} {\bibfnamefont {R.~H.}\ \bibnamefont {Koch}}, \
  and\ \bibinfo {author} {\bibfnamefont {D.~P.}\ \bibnamefont {DiVincenzo}},\
  }\href {\doibase 10.1103/PhysRevB.69.064503} {\bibfield  {journal} {\bibinfo
  {journal} {Phys. Rev. B}\ }\textbf {\bibinfo {volume} {69}},\ \bibinfo
  {pages} {064503} (\bibinfo {year} {2004})}\BibitemShut {NoStop}%
\bibitem [{\citenamefont {Nigg}\ \emph {et~al.}(2012)\citenamefont {Nigg},
  \citenamefont {Paik}, \citenamefont {Vlastakis}, \citenamefont {Kirchmair},
  \citenamefont {Shankar}, \citenamefont {Frunzio}, \citenamefont {Devoret},
  \citenamefont {Schoelkopf},\ and\ \citenamefont {Girvin}}]{Nigg2012}%
  \BibitemOpen
  \bibfield  {author} {\bibinfo {author} {\bibfnamefont {S.~E.}\ \bibnamefont
  {Nigg}}, \bibinfo {author} {\bibfnamefont {H.}~\bibnamefont {Paik}}, \bibinfo
  {author} {\bibfnamefont {B.}~\bibnamefont {Vlastakis}}, \bibinfo {author}
  {\bibfnamefont {G.}~\bibnamefont {Kirchmair}}, \bibinfo {author}
  {\bibfnamefont {S.}~\bibnamefont {Shankar}}, \bibinfo {author} {\bibfnamefont
  {L.}~\bibnamefont {Frunzio}}, \bibinfo {author} {\bibfnamefont {M.~H.}\
  \bibnamefont {Devoret}}, \bibinfo {author} {\bibfnamefont {R.~J.}\
  \bibnamefont {Schoelkopf}}, \ and\ \bibinfo {author} {\bibfnamefont {S.~M.}\
  \bibnamefont {Girvin}},\ }\href {\doibase 10.1103/PhysRevLett.108.240502}
  {\bibfield  {journal} {\bibinfo  {journal} {Phys. Rev. Lett.}\ }\textbf
  {\bibinfo {volume} {108}},\ \bibinfo {pages} {240502} (\bibinfo {year}
  {2012})}\BibitemShut {NoStop}%
\bibitem [{\citenamefont {Ulrich}\ and\ \citenamefont
  {Hassler}(2016)}]{Ulrich2016}%
  \BibitemOpen
  \bibfield  {author} {\bibinfo {author} {\bibfnamefont {J.}~\bibnamefont
  {Ulrich}}\ and\ \bibinfo {author} {\bibfnamefont {F.}~\bibnamefont
  {Hassler}},\ }\href {\doibase 10.1103/PhysRevB.94.094505} {\bibfield
  {journal} {\bibinfo  {journal} {Phys. Rev. B}\ }\textbf {\bibinfo {volume}
  {94}},\ \bibinfo {pages} {094505} (\bibinfo {year} {2016})}\BibitemShut
  {NoStop}%
\bibitem [{\citenamefont {Vool}\ and\ \citenamefont {Devoret}(2017)}]{vool}%
  \BibitemOpen
  \bibfield  {author} {\bibinfo {author} {\bibfnamefont {U.}~\bibnamefont
  {Vool}}\ and\ \bibinfo {author} {\bibfnamefont {M.}~\bibnamefont {Devoret}},\
  }\href {\doibase 10.1002/cta.2359} {\bibfield  {journal} {\bibinfo  {journal}
  {Int. J. Circ. Theor. App.}\ }\textbf {\bibinfo {volume} {45}},\ \bibinfo
  {pages} {897} (\bibinfo {year} {2017})}\BibitemShut {NoStop}%
\bibitem [{\citenamefont {Parra-Rodriguez}\ \emph {et~al.}(2019)\citenamefont
  {Parra-Rodriguez}, \citenamefont {Egusquiza}, \citenamefont {DiVincenzo},\
  and\ \citenamefont {Solano}}]{Parra-Rodriguez2019}%
  \BibitemOpen
  \bibfield  {author} {\bibinfo {author} {\bibfnamefont {A.}~\bibnamefont
  {Parra-Rodriguez}}, \bibinfo {author} {\bibfnamefont {I.~L.}\ \bibnamefont
  {Egusquiza}}, \bibinfo {author} {\bibfnamefont {D.~P.}\ \bibnamefont
  {DiVincenzo}}, \ and\ \bibinfo {author} {\bibfnamefont {E.}~\bibnamefont
  {Solano}},\ }\href {\doibase 10.1103/PhysRevB.99.014514} {\bibfield
  {journal} {\bibinfo  {journal} {Phys. Rev. B}\ }\textbf {\bibinfo {volume}
  {99}},\ \bibinfo {pages} {014514} (\bibinfo {year} {2019})}\BibitemShut
  {NoStop}%
\bibitem [{\citenamefont {Beaudoin}\ \emph {et~al.}(2012)\citenamefont
  {Beaudoin}, \citenamefont {da~Silva}, \citenamefont {Dutton},\ and\
  \citenamefont {Blais}}]{Beaudoin2012}%
  \BibitemOpen
  \bibfield  {author} {\bibinfo {author} {\bibfnamefont {F.}~\bibnamefont
  {Beaudoin}}, \bibinfo {author} {\bibfnamefont {M.~P.}\ \bibnamefont
  {da~Silva}}, \bibinfo {author} {\bibfnamefont {Z.}~\bibnamefont {Dutton}}, \
  and\ \bibinfo {author} {\bibfnamefont {A.}~\bibnamefont {Blais}},\ }\href
  {\doibase 10.1103/PhysRevA.86.022305} {\bibfield  {journal} {\bibinfo
  {journal} {Phys. Rev. A}\ }\textbf {\bibinfo {volume} {86}},\ \bibinfo
  {pages} {022305} (\bibinfo {year} {2012})}\BibitemShut {NoStop}%
\bibitem [{\citenamefont {Strand}\ \emph {et~al.}(2013)\citenamefont {Strand},
  \citenamefont {Ware}, \citenamefont {Beaudoin}, \citenamefont {Ohki},
  \citenamefont {Johnson}, \citenamefont {Blais},\ and\ \citenamefont
  {Plourde}}]{Strand2013}%
  \BibitemOpen
  \bibfield  {author} {\bibinfo {author} {\bibfnamefont {J.~D.}\ \bibnamefont
  {Strand}}, \bibinfo {author} {\bibfnamefont {M.}~\bibnamefont {Ware}},
  \bibinfo {author} {\bibfnamefont {F.}~\bibnamefont {Beaudoin}}, \bibinfo
  {author} {\bibfnamefont {T.~A.}\ \bibnamefont {Ohki}}, \bibinfo {author}
  {\bibfnamefont {B.~R.}\ \bibnamefont {Johnson}}, \bibinfo {author}
  {\bibfnamefont {A.}~\bibnamefont {Blais}}, \ and\ \bibinfo {author}
  {\bibfnamefont {B.~L.~T.}\ \bibnamefont {Plourde}},\ }\href {\doibase
  10.1103/PhysRevB.87.220505} {\bibfield  {journal} {\bibinfo  {journal} {Phys.
  Rev. B}\ }\textbf {\bibinfo {volume} {87}},\ \bibinfo {pages} {220505}
  (\bibinfo {year} {2013})}\BibitemShut {NoStop}%
\bibitem [{\citenamefont {Naik}\ \emph {et~al.}(2017)\citenamefont {Naik},
  \citenamefont {Leung}, \citenamefont {Chakram}, \citenamefont {Groszkowski},
  \citenamefont {Lu}, \citenamefont {Earnest}, \citenamefont {McKay},
  \citenamefont {Koch},\ and\ \citenamefont {Schuster}}]{Naik2017}%
  \BibitemOpen
  \bibfield  {author} {\bibinfo {author} {\bibfnamefont {R.~K.}\ \bibnamefont
  {Naik}}, \bibinfo {author} {\bibfnamefont {N.}~\bibnamefont {Leung}},
  \bibinfo {author} {\bibfnamefont {S.}~\bibnamefont {Chakram}}, \bibinfo
  {author} {\bibfnamefont {P.}~\bibnamefont {Groszkowski}}, \bibinfo {author}
  {\bibfnamefont {Y.}~\bibnamefont {Lu}}, \bibinfo {author} {\bibfnamefont
  {N.}~\bibnamefont {Earnest}}, \bibinfo {author} {\bibfnamefont {D.~C.}\
  \bibnamefont {McKay}}, \bibinfo {author} {\bibfnamefont {J.}~\bibnamefont
  {Koch}}, \ and\ \bibinfo {author} {\bibfnamefont {D.~I.}\ \bibnamefont
  {Schuster}},\ }\href {\doibase 10.1038/s41467-017-02046-6} {\bibfield
  {journal} {\bibinfo  {journal} {Nat. Commun.}\ }\textbf {\bibinfo {volume}
  {8}},\ \bibinfo {pages} {1904} (\bibinfo {year} {2017})}\BibitemShut
  {NoStop}%
\bibitem [{\citenamefont {Reagor}\ \emph {et~al.}(2018)\citenamefont {Reagor},
  \citenamefont {Osborn}, \citenamefont {Tezak}, \citenamefont {Staley},
  \citenamefont {Prawiroatmodjo}, \citenamefont {Scheer}, \citenamefont
  {Alidoust}, \citenamefont {Sete}, \citenamefont {Didier}, \citenamefont
  {da~Silva}, \citenamefont {Acala}, \citenamefont {Angeles}, \citenamefont
  {Bestwick}, \citenamefont {Block}, \citenamefont {Bloom}, \citenamefont
  {Bradley}, \citenamefont {Bui}, \citenamefont {Caldwell}, \citenamefont
  {Capelluto}, \citenamefont {Chilcott} \emph {et~al.}}]{Reagor2018}%
  \BibitemOpen
  \bibfield  {author} {\bibinfo {author} {\bibfnamefont {M.}~\bibnamefont
  {Reagor}}, \bibinfo {author} {\bibfnamefont {C.~B.}\ \bibnamefont {Osborn}},
  \bibinfo {author} {\bibfnamefont {N.}~\bibnamefont {Tezak}}, \bibinfo
  {author} {\bibfnamefont {A.}~\bibnamefont {Staley}}, \bibinfo {author}
  {\bibfnamefont {G.}~\bibnamefont {Prawiroatmodjo}}, \bibinfo {author}
  {\bibfnamefont {M.}~\bibnamefont {Scheer}}, \bibinfo {author} {\bibfnamefont
  {N.}~\bibnamefont {Alidoust}}, \bibinfo {author} {\bibfnamefont {E.~A.}\
  \bibnamefont {Sete}}, \bibinfo {author} {\bibfnamefont {N.}~\bibnamefont
  {Didier}}, \bibinfo {author} {\bibfnamefont {M.~P.}\ \bibnamefont
  {da~Silva}}, \bibinfo {author} {\bibfnamefont {E.}~\bibnamefont {Acala}},
  \bibinfo {author} {\bibfnamefont {J.}~\bibnamefont {Angeles}}, \bibinfo
  {author} {\bibfnamefont {A.}~\bibnamefont {Bestwick}}, \bibinfo {author}
  {\bibfnamefont {M.}~\bibnamefont {Block}}, \bibinfo {author} {\bibfnamefont
  {B.}~\bibnamefont {Bloom}}, \bibinfo {author} {\bibfnamefont
  {A.}~\bibnamefont {Bradley}}, \bibinfo {author} {\bibfnamefont
  {C.}~\bibnamefont {Bui}}, \bibinfo {author} {\bibfnamefont {S.}~\bibnamefont
  {Caldwell}}, \bibinfo {author} {\bibfnamefont {L.}~\bibnamefont {Capelluto}},
  \bibinfo {author} {\bibfnamefont {R.}~\bibnamefont {Chilcott}},  \emph
  {et~al.},\ }\href
  {http://advances.sciencemag.org/lookup/doi/10.1126/sciadv.aao3603} {\bibfield
   {journal} {\bibinfo  {journal} {Sci. Adv.}\ }\textbf {\bibinfo {volume}
  {4}},\ \bibinfo {pages} {eaao3603} (\bibinfo {year} {2018})}\BibitemShut
  {NoStop}%
\bibitem [{\citenamefont {Didier}\ \emph {et~al.}(2018)\citenamefont {Didier},
  \citenamefont {Sete}, \citenamefont {da~Silva},\ and\ \citenamefont
  {Rigetti}}]{Didier2018}%
  \BibitemOpen
  \bibfield  {author} {\bibinfo {author} {\bibfnamefont {N.}~\bibnamefont
  {Didier}}, \bibinfo {author} {\bibfnamefont {E.~A.}\ \bibnamefont {Sete}},
  \bibinfo {author} {\bibfnamefont {M.~P.}\ \bibnamefont {da~Silva}}, \ and\
  \bibinfo {author} {\bibfnamefont {C.}~\bibnamefont {Rigetti}},\ }\href
  {\doibase 10.1103/PhysRevA.97.022330} {\bibfield  {journal} {\bibinfo
  {journal} {Phys. Rev. A}\ }\textbf {\bibinfo {volume} {97}},\ \bibinfo
  {pages} {022330} (\bibinfo {year} {2018})}\BibitemShut {NoStop}%
\bibitem [{\citenamefont {Lu}\ \emph {et~al.}(2017)\citenamefont {Lu},
  \citenamefont {Chakram}, \citenamefont {Leung}, \citenamefont {Earnest},
  \citenamefont {Naik}, \citenamefont {Huang}, \citenamefont {Groszkowski},
  \citenamefont {Kapit}, \citenamefont {Koch},\ and\ \citenamefont
  {Schuster}}]{Lu2017a}%
  \BibitemOpen
  \bibfield  {author} {\bibinfo {author} {\bibfnamefont {Y.}~\bibnamefont
  {Lu}}, \bibinfo {author} {\bibfnamefont {S.}~\bibnamefont {Chakram}},
  \bibinfo {author} {\bibfnamefont {N.}~\bibnamefont {Leung}}, \bibinfo
  {author} {\bibfnamefont {N.}~\bibnamefont {Earnest}}, \bibinfo {author}
  {\bibfnamefont {R.~K.}\ \bibnamefont {Naik}}, \bibinfo {author}
  {\bibfnamefont {Z.}~\bibnamefont {Huang}}, \bibinfo {author} {\bibfnamefont
  {P.}~\bibnamefont {Groszkowski}}, \bibinfo {author} {\bibfnamefont
  {E.}~\bibnamefont {Kapit}}, \bibinfo {author} {\bibfnamefont
  {J.}~\bibnamefont {Koch}}, \ and\ \bibinfo {author} {\bibfnamefont {D.~I.}\
  \bibnamefont {Schuster}},\ }\href {\doibase 10.1103/PhysRevLett.119.150502}
  {\bibfield  {journal} {\bibinfo  {journal} {Phys. Rev. Lett.}\ }\textbf
  {\bibinfo {volume} {119}},\ \bibinfo {pages} {150502} (\bibinfo {year}
  {2017})}\BibitemShut {NoStop}%
\bibitem [{\citenamefont {Huang}\ \emph {et~al.}(2018)\citenamefont {Huang},
  \citenamefont {Lu}, \citenamefont {Kapit}, \citenamefont {Schuster},\ and\
  \citenamefont {Koch}}]{huang2018}%
  \BibitemOpen
  \bibfield  {author} {\bibinfo {author} {\bibfnamefont {Z.}~\bibnamefont
  {Huang}}, \bibinfo {author} {\bibfnamefont {Y.}~\bibnamefont {Lu}}, \bibinfo
  {author} {\bibfnamefont {E.}~\bibnamefont {Kapit}}, \bibinfo {author}
  {\bibfnamefont {D.~I.}\ \bibnamefont {Schuster}}, \ and\ \bibinfo {author}
  {\bibfnamefont {J.}~\bibnamefont {Koch}},\ }\href {\doibase
  10.1103/PhysRevA.97.062345} {\bibfield  {journal} {\bibinfo  {journal} {Phys.
  Rev. A}\ }\textbf {\bibinfo {volume} {97}},\ \bibinfo {pages} {062345}
  (\bibinfo {year} {2018})}\BibitemShut {NoStop}%
\bibitem [{\citenamefont {Wellstood}\ \emph {et~al.}(1987)\citenamefont
  {Wellstood}, \citenamefont {Urbina},\ and\ \citenamefont
  {Clark}}]{Wellstood1987}%
  \BibitemOpen
  \bibfield  {author} {\bibinfo {author} {\bibfnamefont {F.~C.}\ \bibnamefont
  {Wellstood}}, \bibinfo {author} {\bibfnamefont {C.}~\bibnamefont {Urbina}}, \
  and\ \bibinfo {author} {\bibfnamefont {J.}~\bibnamefont {Clark}},\
  }\href@noop {} {\bibfield  {journal} {\bibinfo  {journal} {Appl. Phys.
  Lett.}\ }\textbf {\bibinfo {volume} {50}},\ \bibinfo {pages} {772} (\bibinfo
  {year} {1987})}\BibitemShut {NoStop}%
\bibitem [{\citenamefont {Ithier}\ \emph {et~al.}(2005)\citenamefont {Ithier},
  \citenamefont {Collin}, \citenamefont {Joyez}, \citenamefont {Meeson},
  \citenamefont {Vion}, \citenamefont {Esteve}, \citenamefont {Chiarello},
  \citenamefont {Shnirman}, \citenamefont {Makhlin}, \citenamefont {Schriefl},\
  and\ \citenamefont {Sch{\"{o}}n}}]{Ithier2005}%
  \BibitemOpen
  \bibfield  {author} {\bibinfo {author} {\bibfnamefont {G.}~\bibnamefont
  {Ithier}}, \bibinfo {author} {\bibfnamefont {E.}~\bibnamefont {Collin}},
  \bibinfo {author} {\bibfnamefont {P.}~\bibnamefont {Joyez}}, \bibinfo
  {author} {\bibfnamefont {P.~J.}\ \bibnamefont {Meeson}}, \bibinfo {author}
  {\bibfnamefont {D.}~\bibnamefont {Vion}}, \bibinfo {author} {\bibfnamefont
  {D.}~\bibnamefont {Esteve}}, \bibinfo {author} {\bibfnamefont
  {F.}~\bibnamefont {Chiarello}}, \bibinfo {author} {\bibfnamefont
  {A.}~\bibnamefont {Shnirman}}, \bibinfo {author} {\bibfnamefont
  {Y.}~\bibnamefont {Makhlin}}, \bibinfo {author} {\bibfnamefont
  {J.}~\bibnamefont {Schriefl}}, \ and\ \bibinfo {author} {\bibfnamefont
  {G.}~\bibnamefont {Sch{\"{o}}n}},\ }\href {\doibase
  10.1103/PhysRevB.72.134519} {\bibfield  {journal} {\bibinfo  {journal} {Phys.
  Rev. B}\ }\textbf {\bibinfo {volume} {72}},\ \bibinfo {pages} {134519}
  (\bibinfo {year} {2005})}\BibitemShut {NoStop}%
\bibitem [{\citenamefont {Yoshihara}\ \emph {et~al.}(2006)\citenamefont
  {Yoshihara}, \citenamefont {Harrabi}, \citenamefont {Niskanen}, \citenamefont
  {Nakamura},\ and\ \citenamefont {Tsai}}]{Yoshihara2006}%
  \BibitemOpen
  \bibfield  {author} {\bibinfo {author} {\bibfnamefont {F.}~\bibnamefont
  {Yoshihara}}, \bibinfo {author} {\bibfnamefont {K.}~\bibnamefont {Harrabi}},
  \bibinfo {author} {\bibfnamefont {A.}~\bibnamefont {Niskanen}}, \bibinfo
  {author} {\bibfnamefont {Y.}~\bibnamefont {Nakamura}}, \ and\ \bibinfo
  {author} {\bibfnamefont {J.}~\bibnamefont {Tsai}},\ }\href
  {http://link.aps.org/doi/10.1103/PhysRevLett.97.167001} {\bibfield  {journal}
  {\bibinfo  {journal} {Phys. Rev. Lett.}\ }\textbf {\bibinfo {volume} {97}},\
  \bibinfo {pages} {167001} (\bibinfo {year} {2006})}\BibitemShut {NoStop}%
\bibitem [{\citenamefont {Kumar}\ \emph {et~al.}(2016)\citenamefont {Kumar},
  \citenamefont {Sendelbach}, \citenamefont {Beck}, \citenamefont {Freeland},
  \citenamefont {Wang}, \citenamefont {Wang}, \citenamefont {Yu}, \citenamefont
  {Wu}, \citenamefont {Pappas},\ and\ \citenamefont {McDermott}}]{Kumar2016}%
  \BibitemOpen
  \bibfield  {author} {\bibinfo {author} {\bibfnamefont {P.}~\bibnamefont
  {Kumar}}, \bibinfo {author} {\bibfnamefont {S.}~\bibnamefont {Sendelbach}},
  \bibinfo {author} {\bibfnamefont {M.~A.}\ \bibnamefont {Beck}}, \bibinfo
  {author} {\bibfnamefont {J.~W.}\ \bibnamefont {Freeland}}, \bibinfo {author}
  {\bibfnamefont {Z.}~\bibnamefont {Wang}}, \bibinfo {author} {\bibfnamefont
  {H.}~\bibnamefont {Wang}}, \bibinfo {author} {\bibfnamefont {C.~C.}\
  \bibnamefont {Yu}}, \bibinfo {author} {\bibfnamefont {R.~Q.}\ \bibnamefont
  {Wu}}, \bibinfo {author} {\bibfnamefont {D.~P.}\ \bibnamefont {Pappas}}, \
  and\ \bibinfo {author} {\bibfnamefont {R.}~\bibnamefont {McDermott}},\ }\href
  {\doibase 10.1103/PhysRevApplied.6.041001} {\bibfield  {journal} {\bibinfo
  {journal} {Phys. Rev. Applied}\ }\textbf {\bibinfo {volume} {6}},\ \bibinfo
  {pages} {041001} (\bibinfo {year} {2016})}\BibitemShut {NoStop}%
\bibitem [{\citenamefont {Manucharyan}\ \emph {et~al.}(2009)\citenamefont
  {Manucharyan}, \citenamefont {Koch}, \citenamefont {Glazman},\ and\
  \citenamefont {Devoret}}]{Manucharyan2009}%
  \BibitemOpen
  \bibfield  {author} {\bibinfo {author} {\bibfnamefont {V.~E.}\ \bibnamefont
  {Manucharyan}}, \bibinfo {author} {\bibfnamefont {J.}~\bibnamefont {Koch}},
  \bibinfo {author} {\bibfnamefont {L.~I.}\ \bibnamefont {Glazman}}, \ and\
  \bibinfo {author} {\bibfnamefont {M.~H.}\ \bibnamefont {Devoret}},\ }\href
  {\doibase 10.1126/science.1175552} {\bibfield  {journal} {\bibinfo  {journal}
  {Science}\ }\textbf {\bibinfo {volume} {326}},\ \bibinfo {pages} {113}
  (\bibinfo {year} {2009})}\BibitemShut {NoStop}%
\bibitem [{\citenamefont {Spilla}\ \emph {et~al.}(2015)\citenamefont {Spilla},
  \citenamefont {Hassler}, \citenamefont {Napoli},\ and\ \citenamefont
  {Splettstoesser}}]{Spilla2015}%
  \BibitemOpen
  \bibfield  {author} {\bibinfo {author} {\bibfnamefont {S.}~\bibnamefont
  {Spilla}}, \bibinfo {author} {\bibfnamefont {F.}~\bibnamefont {Hassler}},
  \bibinfo {author} {\bibfnamefont {A.}~\bibnamefont {Napoli}}, \ and\ \bibinfo
  {author} {\bibfnamefont {J.}~\bibnamefont {Splettstoesser}},\ }\href
  {\doibase 10.1088/1367-2630/17/6/065012} {\bibfield  {journal} {\bibinfo
  {journal} {New J. Phys.}\ }\textbf {\bibinfo {volume} {17}},\ \bibinfo
  {pages} {065012} (\bibinfo {year} {2015})}\BibitemShut {NoStop}%
\bibitem [{\citenamefont {Sete}\ \emph {et~al.}(2017)\citenamefont {Sete},
  \citenamefont {Reagor}, \citenamefont {Didier},\ and\ \citenamefont
  {Rigetti}}]{Sete2017}%
  \BibitemOpen
  \bibfield  {author} {\bibinfo {author} {\bibfnamefont {E.~A.}\ \bibnamefont
  {Sete}}, \bibinfo {author} {\bibfnamefont {M.~J.}\ \bibnamefont {Reagor}},
  \bibinfo {author} {\bibfnamefont {N.}~\bibnamefont {Didier}}, \ and\ \bibinfo
  {author} {\bibfnamefont {C.~T.}\ \bibnamefont {Rigetti}},\ }\href {\doibase
  10.1103/PhysRevApplied.8.024004} {\bibfield  {journal} {\bibinfo  {journal}
  {Phys. Rev. Appl.}\ }\textbf {\bibinfo {volume} {8}},\ \bibinfo {pages}
  {024004} (\bibinfo {year} {2017})}\BibitemShut {NoStop}%
\bibitem [{\citenamefont {Diggins}\ \emph {et~al.}(1997)\citenamefont
  {Diggins}, \citenamefont {Whiteman}, \citenamefont {Clark}, \citenamefont
  {Prance}, \citenamefont {Prance}, \citenamefont {Ralph}, \citenamefont
  {Widom},\ and\ \citenamefont {Srivastava}}]{Diggins1997}%
  \BibitemOpen
  \bibfield  {author} {\bibinfo {author} {\bibfnamefont {J.}~\bibnamefont
  {Diggins}}, \bibinfo {author} {\bibfnamefont {R.}~\bibnamefont {Whiteman}},
  \bibinfo {author} {\bibfnamefont {T.}~\bibnamefont {Clark}}, \bibinfo
  {author} {\bibfnamefont {R.}~\bibnamefont {Prance}}, \bibinfo {author}
  {\bibfnamefont {H.}~\bibnamefont {Prance}}, \bibinfo {author} {\bibfnamefont
  {J.}~\bibnamefont {Ralph}}, \bibinfo {author} {\bibfnamefont
  {A.}~\bibnamefont {Widom}}, \ and\ \bibinfo {author} {\bibfnamefont
  {Y.}~\bibnamefont {Srivastava}},\ }\href {\doibase
  10.1016/S0921-4526(96)01152-0} {\bibfield  {journal} {\bibinfo  {journal}
  {Physica B}\ }\textbf {\bibinfo {volume} {233}},\ \bibinfo {pages} {8}
  (\bibinfo {year} {1997})}\BibitemShut {NoStop}%
\bibitem [{\citenamefont {Koch}\ \emph {et~al.}(2009)\citenamefont {Koch},
  \citenamefont {Manucharyan}, \citenamefont {Devoret},\ and\ \citenamefont
  {Glazman}}]{Koch2009}%
  \BibitemOpen
  \bibfield  {author} {\bibinfo {author} {\bibfnamefont {J.}~\bibnamefont
  {Koch}}, \bibinfo {author} {\bibfnamefont {V.}~\bibnamefont {Manucharyan}},
  \bibinfo {author} {\bibfnamefont {M.~H.}\ \bibnamefont {Devoret}}, \ and\
  \bibinfo {author} {\bibfnamefont {L.~I.}\ \bibnamefont {Glazman}},\ }\href
  {\doibase 10.1103/PhysRevLett.103.217004} {\bibfield  {journal} {\bibinfo
  {journal} {Phys. Rev. Lett.}\ }\textbf {\bibinfo {volume} {103}},\ \bibinfo
  {pages} {217004} (\bibinfo {year} {2009})}\BibitemShut {NoStop}%
\bibitem [{\citenamefont {Barone}\ and\ \citenamefont
  {Patern\`o}(1982)}]{barone1982physics}%
  \BibitemOpen
  \bibfield  {author} {\bibinfo {author} {\bibfnamefont {A.}~\bibnamefont
  {Barone}}\ and\ \bibinfo {author} {\bibfnamefont {G.}~\bibnamefont
  {Patern\`o}},\ }\href@noop {} {\emph {\bibinfo {title} {Physics and
  Applications of the Josephson Effect}}}\ (\bibinfo  {publisher} {Wiley},\
  \bibinfo {year} {1982})\ Chap.~\bibinfo {chapter} {12}\BibitemShut {NoStop}%
\bibitem [{\citenamefont {Orlando}\ and\ \citenamefont
  {Delin}(1991)}]{orlando1991foundations}%
  \BibitemOpen
  \bibfield  {author} {\bibinfo {author} {\bibfnamefont {T.~P.}\ \bibnamefont
  {Orlando}}\ and\ \bibinfo {author} {\bibfnamefont {K.~A.}\ \bibnamefont
  {Delin}},\ }\href@noop {} {\emph {\bibinfo {title} {Foundations of Applied
  Superconductivity}}},\ Vol.~\bibinfo {volume} {8}\ (\bibinfo  {publisher}
  {Addison-Wesley Reading, MA},\ \bibinfo {year} {1991})\ Chap.~\bibinfo
  {chapter} {5}\BibitemShut {NoStop}%
\bibitem [{\citenamefont {Brown}(2007)}]{Brown2007}%
  \BibitemOpen
  \bibfield  {author} {\bibinfo {author} {\bibfnamefont {K.~R.}\ \bibnamefont
  {Brown}},\ }\href {\doibase 10.1103/PhysRevA.76.022327} {\bibfield  {journal}
  {\bibinfo  {journal} {Phys. Rev. A}\ }\textbf {\bibinfo {volume} {76}},\
  \bibinfo {pages} {022327} (\bibinfo {year} {2007})}\BibitemShut {NoStop}%
\bibitem [{\citenamefont {Jing}\ \emph {et~al.}(2014)\citenamefont {Jing},
  \citenamefont {Huang},\ and\ \citenamefont {Hu}}]{Jing2014}%
  \BibitemOpen
  \bibfield  {author} {\bibinfo {author} {\bibfnamefont {J.}~\bibnamefont
  {Jing}}, \bibinfo {author} {\bibfnamefont {P.}~\bibnamefont {Huang}}, \ and\
  \bibinfo {author} {\bibfnamefont {X.}~\bibnamefont {Hu}},\ }\href {\doibase
  10.1103/PhysRevA.90.022118} {\bibfield  {journal} {\bibinfo  {journal} {Phys.
  Rev. A}\ }\textbf {\bibinfo {volume} {90}},\ \bibinfo {pages} {022118}
  (\bibinfo {year} {2014})}\BibitemShut {NoStop}%
\bibitem [{\citenamefont {Clerk}\ \emph {et~al.}(2010)\citenamefont {Clerk},
  \citenamefont {Devoret}, \citenamefont {Girvin}, \citenamefont {Marquardt},\
  and\ \citenamefont {Schoelkopf}}]{Clerk2010}%
  \BibitemOpen
  \bibfield  {author} {\bibinfo {author} {\bibfnamefont {A.~A.}\ \bibnamefont
  {Clerk}}, \bibinfo {author} {\bibfnamefont {M.~H.}\ \bibnamefont {Devoret}},
  \bibinfo {author} {\bibfnamefont {S.~M.}\ \bibnamefont {Girvin}}, \bibinfo
  {author} {\bibfnamefont {F.}~\bibnamefont {Marquardt}}, \ and\ \bibinfo
  {author} {\bibfnamefont {R.~J.}\ \bibnamefont {Schoelkopf}},\ }\href
  {\doibase 10.1103/RevModPhys.82.1155} {\bibfield  {journal} {\bibinfo
  {journal} {Rev. Mod. Phys.}\ }\textbf {\bibinfo {volume} {82}},\ \bibinfo
  {pages} {1155} (\bibinfo {year} {2010})}\BibitemShut {NoStop}%
\bibitem [{\citenamefont {Qiu}\ \emph {et~al.}(2016)\citenamefont {Qiu},
  \citenamefont {Xiong}, \citenamefont {He}, \citenamefont {Li},\ and\
  \citenamefont {You}}]{Qiu2016}%
  \BibitemOpen
  \bibfield  {author} {\bibinfo {author} {\bibfnamefont {Y.}~\bibnamefont
  {Qiu}}, \bibinfo {author} {\bibfnamefont {W.}~\bibnamefont {Xiong}}, \bibinfo
  {author} {\bibfnamefont {X.-L.}\ \bibnamefont {He}}, \bibinfo {author}
  {\bibfnamefont {T.-F.}\ \bibnamefont {Li}}, \ and\ \bibinfo {author}
  {\bibfnamefont {J.~Q.}\ \bibnamefont {You}},\ }\href {\doibase
  10.1038/srep28622} {\bibfield  {journal} {\bibinfo  {journal} {Sci. Rep.}\
  }\textbf {\bibinfo {volume} {6}},\ \bibinfo {pages} {28622} (\bibinfo {year}
  {2016})}\BibitemShut {NoStop}%
\bibitem [{\citenamefont {Groszkowski}\ \emph {et~al.}(2018)\citenamefont
  {Groszkowski}, \citenamefont {Paolo}, \citenamefont {Grimsmo}, \citenamefont
  {Blais}, \citenamefont {Schuster}, \citenamefont {Houck},\ and\ \citenamefont
  {Koch}}]{Groszkowski2017b}%
  \BibitemOpen
  \bibfield  {author} {\bibinfo {author} {\bibfnamefont {P.}~\bibnamefont
  {Groszkowski}}, \bibinfo {author} {\bibfnamefont {A.~D.}\ \bibnamefont
  {Paolo}}, \bibinfo {author} {\bibfnamefont {A.~L.}\ \bibnamefont {Grimsmo}},
  \bibinfo {author} {\bibfnamefont {A.}~\bibnamefont {Blais}}, \bibinfo
  {author} {\bibfnamefont {D.~I.}\ \bibnamefont {Schuster}}, \bibinfo {author}
  {\bibfnamefont {A.~A.}\ \bibnamefont {Houck}}, \ and\ \bibinfo {author}
  {\bibfnamefont {J.}~\bibnamefont {Koch}},\ }\href {\doibase
  10.1088/1367-2630/aab7cd} {\bibfield  {journal} {\bibinfo  {journal} {New J.
  Phys.}\ }\textbf {\bibinfo {volume} {20}},\ \bibinfo {pages} {043053}
  (\bibinfo {year} {2018})}\BibitemShut {NoStop}%
\end{thebibliography}%

\end{document}